# Is the brain macroscopically linear?
# A system identification of resting state dynamics

Erfan Nozari[1], Maxwell A. Bertolero[2], Jennifer Stiso[2,3], Lorenzo Caciagli[2],
Eli J. Cornblath[2,3], Xiaosong He[2], Arun S. Mahadevan[2], George J. Pappas[4], and
Dani Smith Bassett[2,4,5,6,7,8,*]

[1]*Department of Mechanical Engineering, University of California, Riverside, CA, USA*
[2]*Department of Bioengineering, University of Pennsylvania, Philadelphia, PA, USA*
[3]*Department of Neuroscience, University of Pennsylvania, Philadelphia, PA, USA*
[4]*Department of Electrical and Systems Engineering, University of Pennsylvania, Philadelphia, PA, USA*
[5]*Department of Physics & Astronomy, University of Pennsylvania, Philadelphia, PA, USA*
[6]*Department of Neurology, University of Pennsylvania, Philadelphia, PA, USA*
[7]*Department of Psychiatry, University of Pennsylvania, Philadelphia, PA, USA*
[8]*Santa Fe Institute, Santa Fe, NM, USA*
[*]*Corresponding author (email: dsb@seas.upenn.edu)*

## Abstract

A central challenge in the computational modeling of neural dynamics is the trade-off between accuracy and simplicity. At the level of individual neurons, nonlinear dynamics are both experimentally established and essential for neuronal functioning. One may therefore expect the collective dynamics of massive networks of such neurons to exhibit an even larger repertoire of nonlinear behaviors. An implicit assumption has thus formed that an "accurate" computational model of whole-brain dynamics must inevitably be nonlinear whereas linear models may provide a first-order approximation. To what extent this assumption holds, however, has remained an open question. Here, we provide new evidence that challenges this assumption at the level of whole-brain blood-oxygen-level-dependent (BOLD) and macroscopic field potential dynamics by leveraging the theory of system identification. Using functional magnetic resonance imaging (fMRI) and intracranial electroencephalography (iEEG), we model the spontaneous, resting state activity of 700 subjects in the Human Connectome Project (HCP) and 122 subjects from the Restoring Active Memory (RAM) project using state-of-the-art linear and nonlinear model families. We assess relative model fit using predictive power, computational complexity, and the extent of residual dynamics unexplained by the model. Contrary to our expectations, linear auto-regressive models achieve the best measures across all three metrics. To understand and explain this linearity, we highlight four properties of macroscopic neurodynamics which can counteract or mask microscopic nonlinear dynamics: averaging over space, averaging over time, observation noise, and limited data samples. Whereas the latter two are technological limitations and can improve in the future, the former two are inherent to aggregated macroscopic brain activity. Our results demonstrate the discounted potential of linear models in accurately capturing macroscopic brain dynamics. This, together with the unparalleled interpretability of linear models, can greatly facilitate our understanding of macroscopic neural dynamics, which in turn may facilitate the principled design of model-based interventions for the treatment of neuropsychiatric disorders.

Throughout the recent history of neuroscience, computational models have been developed and used ubiquitously in order to decompose the complex neural mechanisms underlying cognition and behavior [1–3]. A dilemma that is inherent to computational modeling, but particularly challenging in computational neuroscience, is the trade-off between (cross-validated) accuracy and simplicity. Both finely detailed models [4] and broadly simplified ones [5, 6] have their respective proponents. One of the many facets of this trade-off pertains to the use of linear *versus* nonlinear models. Nonlinearity of dynamics is inevitable at the (micro) scale of individual neurons [7] and their components [8], and has been demonstrated, though less comprehensively, at the (meso) scale of neuronal populations [9]. Further supported by theoretical derivations [10] and motivated by the much larger repertoire of behaviors of nonlinear systems (including chaos, multistability, and meta-stability), an assumption has thus formed [11–14] that accurate models of neurodynamics at the macroscale (of brain regions) must inevitably be nonlinear.

This assumption begs the question of whether nonlinear models will in fact perform better than linear ones in accounting for the dynamics of neuroimaging and neurophysiological data. Specifically, can nonlinear models explain neuroimaging or neurophysiological data more accurately than linear ones? This pragmatic modeling question, importantly, is different from the general question of whether any signs of "nonlinearity" can be found in neuroimaging time series [15–17] (see Discussion). Few investigations [18–20] have indeed sought to answer the former question directly by comparing the "fit" of linear and nonlinear models to neurophysiological (EEG and iEEG) time series. Even these few works are limited, however, in that each provides a single comparison between a linear and a nonlinear family of models, which need not be the best representatives of linear and nonlinear models in general. These works have also arrived at inconsistent conclusions, providing little resolution as to the ultimate choice of models to be implemented in computational neuroscience.

In this work, we first provide a side-by-side comparison of various state-of-the-art families of linear and nonlinear ordinary differential equation (ODE) models of neurodynamics in terms of their cross-validated fit to resting state fMRI time series of 700 subjects from the Human Connectome Project (HCP) and resting state iEEG time series from 122 subjects from the Restoring Active Memory (RAM) project (see Methods). Our modeling and model comparison approaches are based on the theory of system identification [21], a core discipline in systems and control theory concerned with the data-driven model construction and evaluation of dynamical systems. Motivated by the prediction error framework in system identification theory, we use three metrics for comparison, namely, the normalized sum of squared prediction errors, the whiteness of residuals, and the computational complexity of each method. These metrics measure, respectively, how accurately the model can predict future values of the time series from its past, how much unmodeled dynamics have remained in the model fitting residuals, and how much CPU time it takes to fit the model and cross-validate its prediction accuracy. Across all measures, we find that linear models outperform nonlinear ones, thereby simultaneously maximizing accuracy and simplicity (both as quantified by our aforementioned measures).

In the second part of the paper, we seek to answer the question of why nonlinear models do not provide more accurate predictions than linear ones despite the fact that neurodynamics are inevitably nonlinear at the microscale. Specifically, we numerically demonstrate, using a simple sigmoidal nonlinearity, that four properties of macroscopic brain dynamics can fundamentally *counteract* or apparently *mask* nonlinear dynamics present at the microscale: averaging over the activity of large populations of neurons to obtain a single macroscopic time series (averaging over space), natural low pass-filtering properties of brain processes (averaging over time), observation noise, and limited data samples. While the effects of observation noise and limited data samples are technology dependent but otherwise independent from the form of nonlinearity, the effects of spatiotemporal averaging are fundamental to macroscopic neural dynamics, and may depend on the functional form of the microscale nonlinearity. We thus also verify and demonstrate the effects of spatiotemporal averaging using a data-driven and biophysically grounded spiking neuron model [22]. Together, our results provide important evidence against the common presumption of nonlinearity in computational neuroscience, as well as a methodology based on system identification theory to quantitatively define a "best" model of whole brain dynamics given a set of specified costs.

## Results

**System identification and data-driven computational modeling.** Among the diverse categories of computational models used in neuroscience, we fo-cus on ODE models of the general form

$$\dot{\mathbf{x}}(t) = f(\mathbf{x}(t)) + \mathbf{e}_1(t), \qquad \mathbf{x}(0) = \mathbf{x}_0 \quad (1a)$$
$$\mathbf{y}(t) = h(\mathbf{x}(t)) + \mathbf{e}_2(t) \quad (1b)$$

where $\mathbf{y}(t)$ is an $n$-dimensional time series of recorded brain activity, in this case via resting state fMRI (rsfMRI) or iEEG (rsiEEG), $\mathbf{x}(t)$ is an $m$-dimensional time series of "internal" brain states, $f$ and $h$ are generally nonlinear vector fields, and $\mathbf{e}_1(t)$ and $\mathbf{e}_2(t)$ are time series of process and measurement noise with arbitrary statistics (Figure 1a-b). It is possible, although not necessary, that $m = n$. As with any differential equation, the description would not be complete without the initial condition $\mathbf{x}(0) = \mathbf{x}_0$, determining the state of the brain at the initial recording time $t = 0$. Note that no external input $\mathbf{u}(t)$ is considered due to the resting state condition of the experiments. Also, given that we can only sample $\mathbf{y}(t)$ in discrete time, we implement Eq. (1) by approximating the derivative $\dot{\mathbf{x}}(t)$ as a first difference $\mathbf{x}(t) - \mathbf{x}(t-1)$ (see Methods).

Throughout the field of computational neuroscience, numerous models of the form in Eq. (1) or its discretization (see Methods) are constructed and used, each with different functional forms and noise statistics [4, 5, 7, 10, 13, 20, 23]. The critical but fairly overlooked quest of system identification [21] is then to find the "best" model, among all the available options, against experimental data. This comparison, indeed, depends on one's measure of a model's goodness of fit.

A natural choice, referred to as the prediction error (PE) approach, is based on how well a given model can predict the *future* of the measured time series from its *current and past* values (Figure 1c). Note that this prediction is precisely what an ODE such as Eq. (1a) defines: it models the *change* $\dot{\mathbf{x}}(t)$, and thus the immediate future, of the system's state (and therefore output) from its current state $\mathbf{x}(t)$. Since the state $\mathbf{x}(t)$ is not directly available, it should in turn be estimated from the current and past measurements of the output $\mathbf{y}(t)$. Therefore, the PE approach in its simplest form seeks to minimize, within any given parametric or non-parametric family of models, the one-step ahead PE

$$\varepsilon(t) = \mathbf{y}(t) - \hat{\mathbf{y}}(t|t-1) \quad (2)$$

where $\hat{\mathbf{y}}(t|t-1)$ is the Bayes-optimal, minimum variance estimate of $\mathbf{y}(t)$ given all of the history $\{\mathbf{y}(0), \ldots, \mathbf{y}(t-1)\}$ of $\mathbf{y}$ up to time $t-1$ [21] (Figure 1c). Notably, the PE approach focuses on the prediction accuracy of the time series itself, rather than the prediction accuracy of functional connectivity (Figure 1d) or other statistics of the time series (cf. Discussion and Supplementary Figure 1).

The task of system identification does not end once the parameters of a given family of models are fit to the (training) data. The critical next step is to assess the quality of the fit, particularly to data withheld during the training (cross-validation). In the PE approach, the two most widely used measures are the variance and the whiteness of the PE [21], where the former is often measured by

$$R_i^2 = 1 - \frac{\sum_t \varepsilon_i(t)^2}{\sum_t (y_i(t) - \bar{y}_i)^2} \quad (3)$$

and the latter is often assessed via a $\chi^2$ test of whiteness (see Methods), for each channel $i = 1, \ldots, n$. In Eq. (3), $\varepsilon(t)$ is the same one-step-ahead prediction error in Eq. (2), and $\bar{y}_i$ is the temporal average of $y_i(t)$ (often equal to zero due to mean centering) and corresponds to a constant predictor which always predicts $\mathbf{y}(t)$ equal to its average $\bar{\mathbf{y}}$. Therefore, it is clear that $R_i^2$ is always less than or equal to 1 but *can be negative*. A value of $R_i^2 = 1$ indicates a perfect model (for channel/region $i$), $R_i^2 = 0$ indicates a model as good as the constant predictor, and $R_i^2 < 0$ indicates a model worse than the constant predictor.

**Linear models: maximum prediction accuracy with minimum computational complexity.** In this work, we fit and compare several families of linear and nonlinear models, as described below (see Methods for details). We fit each family of models to the data for each subject, thereby finding the optimal model at the global or local level (if the corresponding optimization algorithm is convex or non-convex, respectively). We then compare the resulting best models in each family in terms of their cross-validated fit to held-out data of the same subject (see Methods). The most important ground for comparison is the accuracy of their fit, measured by $R_i^2$ according to Eq. (3).

First, consider the results for the fMRI data (Figure 2a-b). While we describe the results obtained using a relatively coarse parcellation here, similar results also hold for finely-parcellated and unparcellated data (cf. Supplementary Figures 11-13). Over-all, linear models that directly fit the BOLD time series without (de)convolving with an HRF, either with dense or sparse effective connectivity, and with or without higher-order auto-regressive lags, achieve the highest $R^2$. Among nonlinear models, the manifold-based locally-linear model achieves a comparable $R^2$. Yet, upon closer inspection of this model, we observe that its window size (which is chosen optimally, see Methods and Supplementary Figure 9) is very large, effectively making it a globally-linear model. The lack of nonlinearity becomes even clearer when examining the pairwise models. Here, we see that a simple linear model performs as well as the MMSE model, or even slightly better (Figure 2b, right panel) due to the numerical errors of distribution estimation. We thus infer that the former achieves the highest prediction accuracy achievable by *any* generally nonlinear model, albeit for pairwise prediction.

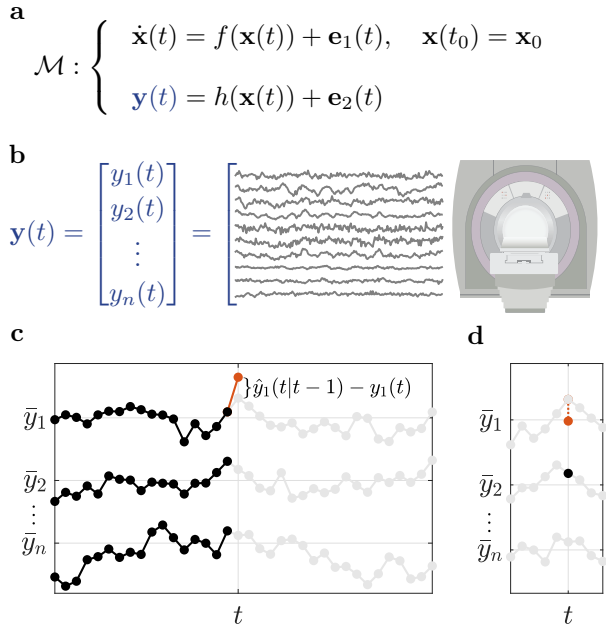

**Figure 1: Prediction error method for system identification.** (a) The general category of computational models $\mathcal{M}$ studied in this work, represented by an ODE describing the resting state evolution of internal states $\mathbf{x}(t)$ and an output model that maps internal states to fMRI/iEEG time series $\mathbf{y}(t)$, as shown for fMRI in panel (b). A total of $n_{\text{fMRI}} = 116$ regions are used throughout (see Methods) for fMRI, while $13 \le n_{\text{iEEG}} \le 175$ channels are used for each iEEG subject. (c) A schematic representation of the prediction error system identification framework used in this work. At each time $t$, all of $\{\mathbf{y}(0), \ldots, \mathbf{y}(t-1)\}$ is used to predict $\mathbf{y}(t)$, simultaneously across all channels, which is denoted for simplicity by $\hat{\mathbf{y}}(t|t-1)$. $\bar{y}_i$ denotes the temporal average of $y_i(t)$ for each channel $i$. This estimate should not be confused with estimates of functional connectivity (FC). (d) FC measures the covariation between pairs of channels or, equivalently, how well each $y_i(t)$ ($y_1(t)$ in the figure) can be predicted from each other $y_j(t)$ ($y_2(t)$ in the figure), *at the same time t*.

The second ground for comparison is the whiteness of model residuals, also in held-out data, which indicates that all the dynamics in the data are captured by the model and have not leaked into the residuals (Figure 2c-d). Here, linear models also score higher than nonlinear models; all except the subspace method have median $p$-values above 0.05, indicating that their residuals are white enough for the null hypothesis of whiteness not to be rejected ($\chi^2$ test of whiteness, see Methods). The autoregressive (AR) models clearly outperform the others. Generally, the number of lags and sparsity patterns have little effect on the prediction accuracy of linear AR models for rsfMRI data, a positive but weak effect on the whiteness of the residual, and a negative effect on the computational complexity (Supplementary Figure 6). Similar to the comparison of $R^2$ values, the only nonlinear model whose whiteness of residuals is comparable to the linear ones is the manifold-based locally-linear model which, as explained above, is effectively linear at the global scale. Also as before, the pairwise linear models achieve a degree of whiteness that is almost identical to the pairwise MMSE estimator, ensuring their optimality among all linear and nonlinear pairwise predictors.

Third and finally, we can compare the models by considering the total time that it takes for their learning and prediction (Figure 2e). When comparing the most efficient linear and nonlinear models, we find that linear models take at least one order of magnitude less time to fit than nonlinear models, as ex-pected. However, linear methods can also be extremely complex to learn; linear models with states at the neural level ('Linear w/ HRF') require the most time to learn due to their high flexibility. Notably, this additional complexity of the 'Linear w/ HRF' or nonlinear methods is not counterbalanced by any benefits in their accuracy or whiteness of residuals, making the simplest linear models the preferred choice across all measures.

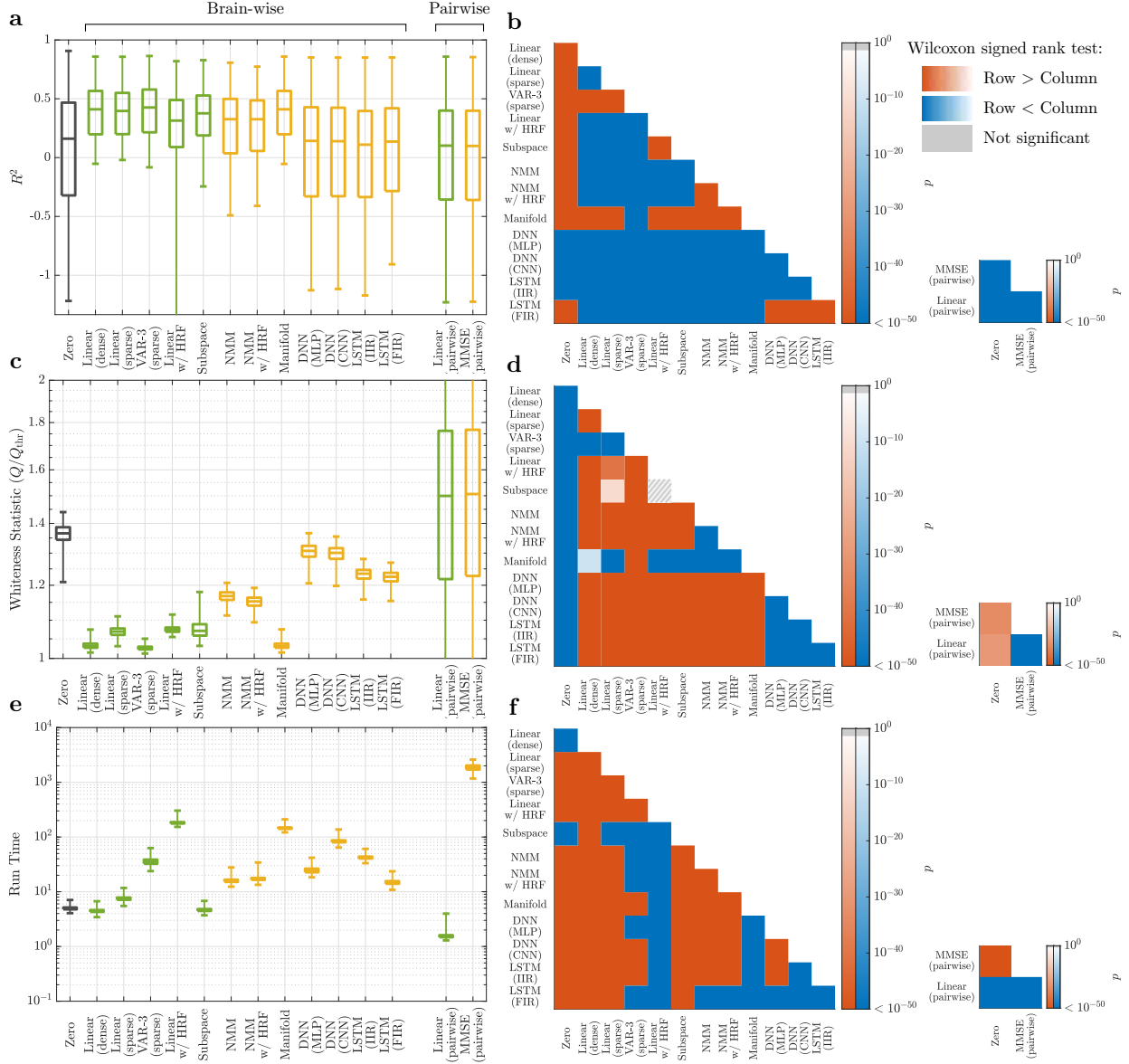

Figure 2: **Linear vs. nonlinear models of rsfMRI activity.** (**a**) The distribution of cross-validated regional $R_i^2$, combined across all 116 regions and 700 subjects (sample size = 81200), for linear (green) and nonlinear (yellow) models. The gray box corresponds to the zero model used as a baseline (see Methods for an explanation of each model). The 10 methods on the left are brain-wise in the sense that they use information from all regions simultaneously to predict each region's next time point; in contrast, the two methods on the right are pairwise, predicting the next time point of each region based on the history of every other region (including itself) alone. The boxplots of the latter two methods thus contain $116^2 \times 700$ samples. Negative values of $R^2$ indicate that models have a worse prediction performance than a constant predictor which always predicts the next value of a signal to be equal to its mean. (**b**) The $p$-value of one-sided Wilcoxon signed rank test performed between all pairs of brain-wise distributions (left) and all pairs of pairwise distributions (right) of $R^2$ in panel (**a**). Warm (cold) colors indicate that the distribution labelled on the row has significantly larger (smaller) samples than the distribution labelled on the column. Gray hatches indicate non-significant differences at an $\alpha = 0.05$ with BH-FDR correction for multiple comparisons. (**c, d**) Similar to panels (**a, b**) but for the statistic $Q$ of the multivariate test of whiteness relative to its rejection threshold $Q_{\text{thr}}$ (cf. Methods). Smaller $Q/Q_{\text{thr}}$ indicates whiter (better) residuals, with $Q/Q_{\text{thr}} \leq 1$ required for the null hypothesis of whiteness not to be rejected. (**e, f**) Similar to panels (**a, b**) except for the time that it took for the learning and out of sample prediction of each model to run, per subject per cross validation (see Methods). In all box plots, the center line, box limits, and whiskers represent the median, upper and lower quartiles, and the smallest and largest samples, respectively.

Next, we perform the same comparisons between linear and nonlinear models, but now on the basis of their fit to resting state *iEEG* field potential dynamics (Figure 3). Similar to rsfMRI data, linear AR models provide the best fit to the data, in terms of both the magnitude and whiteness of their cross-validated prediction error. These models also have lower computational complexity than nonlinear ones, with about an order of magnitude (or higher) advantage in computation time.

Alongside these similarities between the rsfMRI and rsiEEG data, two major distinctions are notable. First, the $R^2$ values are generally much higher for iEEG, as evidenced by the $R^2$ distributions of the zero model between the two cases. This difference is due to the fact that the iEEG time series has a much higher sampling rate and is therefore smoother. As a result, even predicting each sample equal to its previous sample (i.e., the zero model) has a median accuracy of more than 97% (see Supplementary Figures 14-16 for a more detailed assessment of the effects of sampling rate on models' $R^2$). This fact only highlights the importance of the zero model; without it, the $R^2$ of all models might have seemed satisfactorily high. In comparison to the zero model, however, it becomes clear that a simple 1-lag linear model, for example, has in fact a very low predictive power.

The second major distinction between the two modalities is the amount of history and temporal dependency within them. fMRI data is almost Markovian, so that $\mathbf{y}(t-1)$ contains almost all the information available for the prediction of $\mathbf{y}(t)$. Little information is also contained in $\mathbf{y}(t-2)$, but almost no information is contained in time points further in the past (Figure 2a). When considering iEEG data, in contrast, increasing the number of autoregressive lags up to about 100 still improves the $R^2$, although the exact optimal number of lags varies between data segments. In this comparison, it is also important to take into account the vast difference in the sampling frequencies between the modalities, where 2 lags in the fMRI dataset amounts to 1.44 seconds while 100 lags of the ECoG data sums to only 0.2 seconds. This greater *richness* of iEEG dynamics from a modeling perspective is also responsible, at least in part, for the markedly lower whiteness of residuals of all model families with respect to fMRI (see Figure 3c vs. Figure 2c). This greater richness of iEEG is also consistent with, though not necessarily a direct consequence of, the fact that iEEG data reflects neural signals more directly than fMRI.

**Why linear? The linearizing effects of macroscopic neurodynamics and neuroimaging.** The above results pose the natural question of why nonlinear models were not able to capture the dynamics in rs-fMRI/rs-iEEG data beyond linear ones, even though microscopic neuronal dynamics are fundamentally nonlinear. Here, we focus on four properties of macroscopic neurodynamics and neuroimaging, and show that, in principle, they either fundamentally counteract or apparently mask nonlinearities. Due to its unique position in neural modeling [9, 10], we will use the sigmoidal nonlinearity to illustrate these effects; we note, however, that the effects are otherwise applicable to other forms of nonlinearity.

The first property that can fundamentally *counteract* microscopic nonlinearities is spatial averaging. Imaging tools that are capable of measuring macroscopic brain dynamics detect a signal that reflects an average over the activity of hundreds, thousands, or even millions of neurons. This spatial averaging can weaken, rather quickly, the nonlinear relationships in the dynamics of individual units (neurons or small-scale neuronal populations) as long as the units are not perfectly correlated, and can completely nullify nonlinearities when correlations decay with distance (Figure 4a-c). Note that this distance can be the physical distance between the units, as assumed here, or in any relevant space such as that of neural codes and stimulus preference. The key factor in the linearizing effect of spatial averaging is the decay of pairwise correlations between neurons so that not all pairs of neurons in a region are significantly correlated (a state of blanket global synchrony).

This linearizing effect of spatial averaging is similar to, but different from, stochastic linearization (a.k.a., quasi-linearization) [24]. While the latter *approxi-*

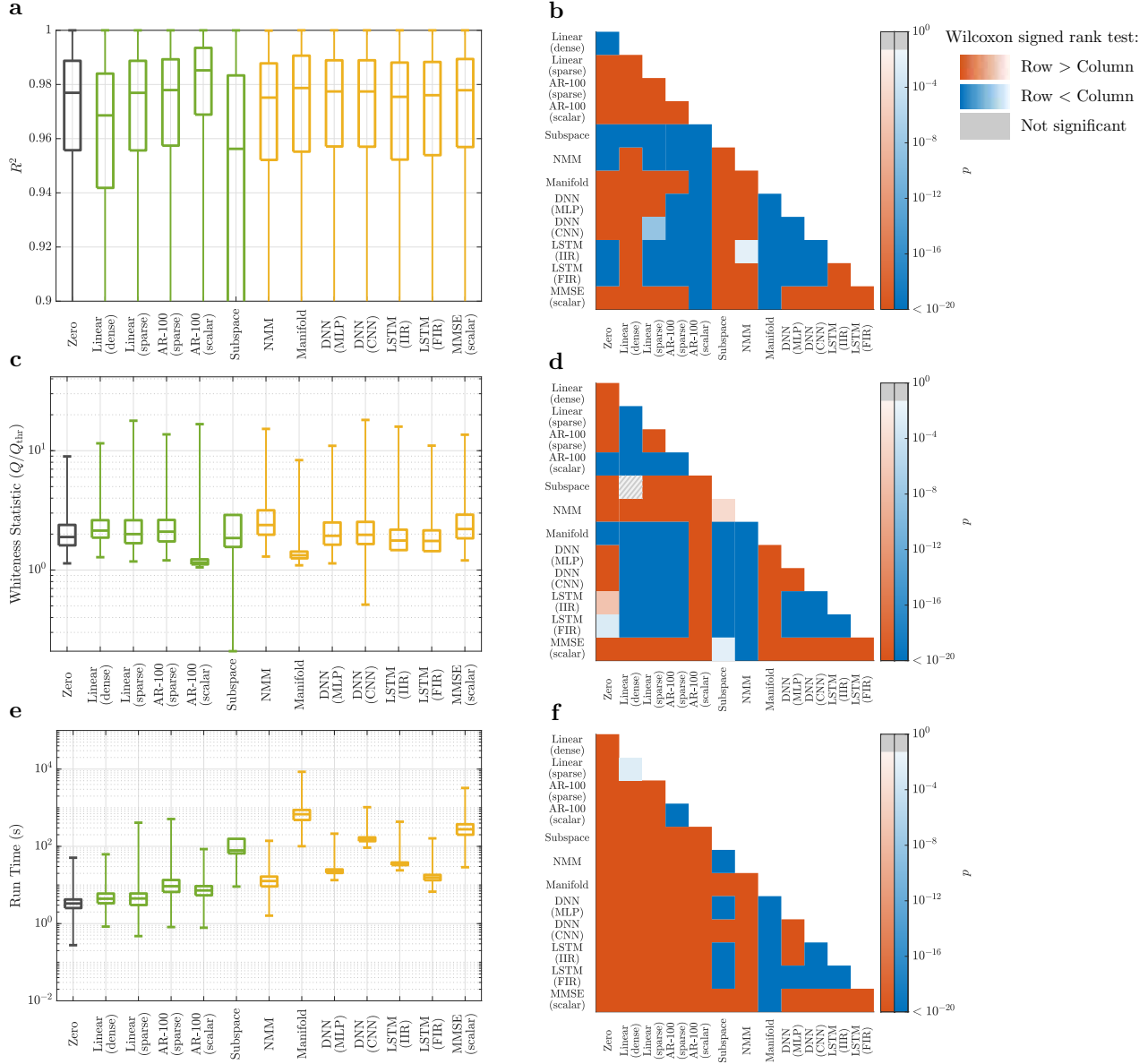

**Figure 3: Linear versus nonlinear models of rsiEEG activity.** Panels parallel those in Figure 2. **(a)** The distribution of cross-validated regional $R_i^2$, combined across all electrodes (the number of which varies among subjects) and all the recording sessions of the 122 subjects (sample size = 776484). Linear and nonlinear methods are depicted by green and yellow boxes, respectively (see Methods for an explanation of each model). Unlike data presented in Figure 2, pairwise linear or pairwise MMSE models are not included due to the observation that between-electrode connections decrease the cross-validated accuracy of the top model (cf. the 4th and 5th box plots). In contrast, including scalar autoregressive lags is highly beneficial in iEEG, whereas it is not so in rsfMRI. Therefore, the MMSE model here is scalar, conditioning on the past lags of each region itself. The lower whisker of the box plots are trimmed to allow for better illustration of the interquartile ranges. **(b)** The $p$-value of the one-sided Wilcoxon signed rank test performed between all pairs of distributions of $R^2$ in panel **(a)**. Warm (cold) colors indicate that the distribution labelled on the row is significantly larger (smaller) than the distribution labelled on the column. Gray hatches indicate non-significant differences evaluated at $\alpha = 0.05$ with BH-FDR correction for multiple comparisons. **(c, d)** Similar to panels **(a, b)** but for statistic $Q$ of the multivariate test of whiteness relative to its rejection threshold $Q_{\text{thr}}$ (cf. Methods). Smaller $Q/Q_{\text{thr}}$ indicates whiter (better) residuals, with $Q/Q_{\text{thr}} \leq 1$ required for the null hypothesis of whiteness not to be rejected. **(e, f)** Similar to panels **(a, b)** but for the time that it took for the learning and out-of-sample-prediction of each model. In all box plots, the center line, box limits, and whiskers represent the median, upper and lower quartiles, and the smallest and largest samples, respectively.

*mates* the relationship $y = \sigma(x)$ using its expected slope $E[\partial y/\partial x]$, spatial averaging as discussed here can result in a relationship that is truly linear. Also, the same effect can be observed when averaging other forms of nonlinearity than the sigmoid. Supplementary Figure 2 shows the effect of spatial averaging on spiking neurons evolving according to the Izhikevic model [22]. This model has completely different nonlinearities than the sigmoid (polynomial and discontinuous) and shows a robust nonlinear phenomenon (limit cycle). Although more than a few (but still no more than 100-$10^4$) neurons are required, spatial averaging still dissolves the nonlinear aspects of the dynamics, while mostly sparing the linear ones.

The second property capable of completely counteracting microscale nonlinearities is temporal averaging. Macroscopic neural dynamics are often observed, or even defined, through signals that are low-pass filtered versions of micro- and meso-scale variables. The most notable of these is perhaps the BOLD signal captured by fMRI, which can be seen as an observation of neural activity passed through the low-pass filter of the HRF. Similarly, although to a lesser extent, the local field potentials captured by iEEG most strongly reflect the aggregate pyramidal post-synaptic currents [25], which are themselves low-pass filtered observations of spiking activity through synaptic transmission and neuronal membranes' resistive-capacitive circuit. The effect of low-pass filtering, in essence, is temporal averaging, which impacts nonlinearities in a manner that is similar to that of spatial averaging (Figure 4e-f). The parallel of spatial correlations here is the autocorrelation function or its frequency-domain representation, the power spectral density (PSD). Autocorrelation represents how the correlation between adjacent samples of a signal decay with the temporal distance between those samples. As expected, the smaller the bandwidth of the signal (i.e., the faster their PSD decays with frequency before low-pass filtering), the weaker the linearizing effect of low-pass filtering. As a result, stronger low-pass filtering would also be required to completely nullify nonlinear relationships in signals with narrower bandwidth (Figure 4d). The linearizing effect of temporal averaging also holds for deterministic dynamics, albeit with the resulting linear dynamics (post averaging) also being deterministic (Supplementary Figure 3).

A third property that can counteract or mask nonlinearities is noise. Although both process noise and observation (scanner or electrode) noise may have linearizing effects, here we focus only on the latter. As with any neuroimaging time series, various sources of observation noise can affect the fMRI/iEEG time series [26, 27] and, in turn, "blur" nonlinear relationships, even if they exist between the underlying noise-free BOLD/LFP time series (Figure 4g-h). In fact, when the power of noise reaches the power of the signal (SNR $\sim$ 1), it can completely mask a nonlinear relationship in the absence of any spatial or temporal averaging. In reality, however, the linearizing effect of observation noise can combine with spatiotemporal averaging, making the $2 \lesssim \text{SNR} \lesssim 14$ that we have in rsfMRI data (Supplementary Figure 8) potentially more than enough to mask any remaining nonlinearities post-spatiotemporal averaging. Ironically, the use of linear filtering to "clean the data" is more likely to further linearize the dynamics of the time series due to temporal averaging effects discussed above, instead of recovering nonlinearities lost due to noise (Supplementary Note 1). Nonlinear post-processing steps, on the other hand, may leave their own potentially nonlinear signatures in the data, but such signatures should not be confused with true nonlinear relationships in the original BOLD/LFP signal. Further, although we let the noise in Figure 4g-h be independent of the signal, as is typically the case for measurement noise, this linearizing effect would still hold if the noise is linearly dependent on the signal.

The fourth and final property that we will discuss is the number of samples required for detecting nonlinear relationships in large dimensions. Let us assume, despite our discussion so far, that a perfect noise-free nonlinear relationship exists between $n$-dimensional fMRI or iEEG time series and a noise-free sensor can capture it perfectly. When only $N \sim 1000$ data points are available, we find that the manifold-based predictor – which was our most predictive nonlinear method both for fMRI and iEEG – is still unable to predict the nonlinear relationship better than a linear model in $n \sim 40$ dimensions or higher (Figure 4i-j). This loss in the predictive power of this nonlinear predictor with increasing dimensionality can be easily seen from the fact that the smallest mesh, having two points per dimension, requires an exponentially large $N = 2^n$ data points. Indeed, incorporating structural bias into the learning algorithm can arbitrarily reduce this sample complexity *if* the incorporated bias is consistent with the underlying data [28] (e.g., if one looks for relationships of the form $y = \sigma(x_1 + \cdots + x_n)$ in Figure 4i-j). However, using predictors with structural bias can also be arbitrarily misleading if their form of nonlinearity is not consistent with the given data [29], which is one potential reason for the lower performance of most nonlinear methods in Figures 2 and 3. This discussion also makes it clear that the inability of our nonlinear system identification methods to outperform linear ones in Figures 2 and 3 over the entire brain is not a proof that no nonlinear method can possibly do so. We can, nevertheless, be certain about this for pairwise or scalar AR models (for fMRI and iEEG, respectively) where the optimal MMSE predictor was computable and performed as well as a linear one.

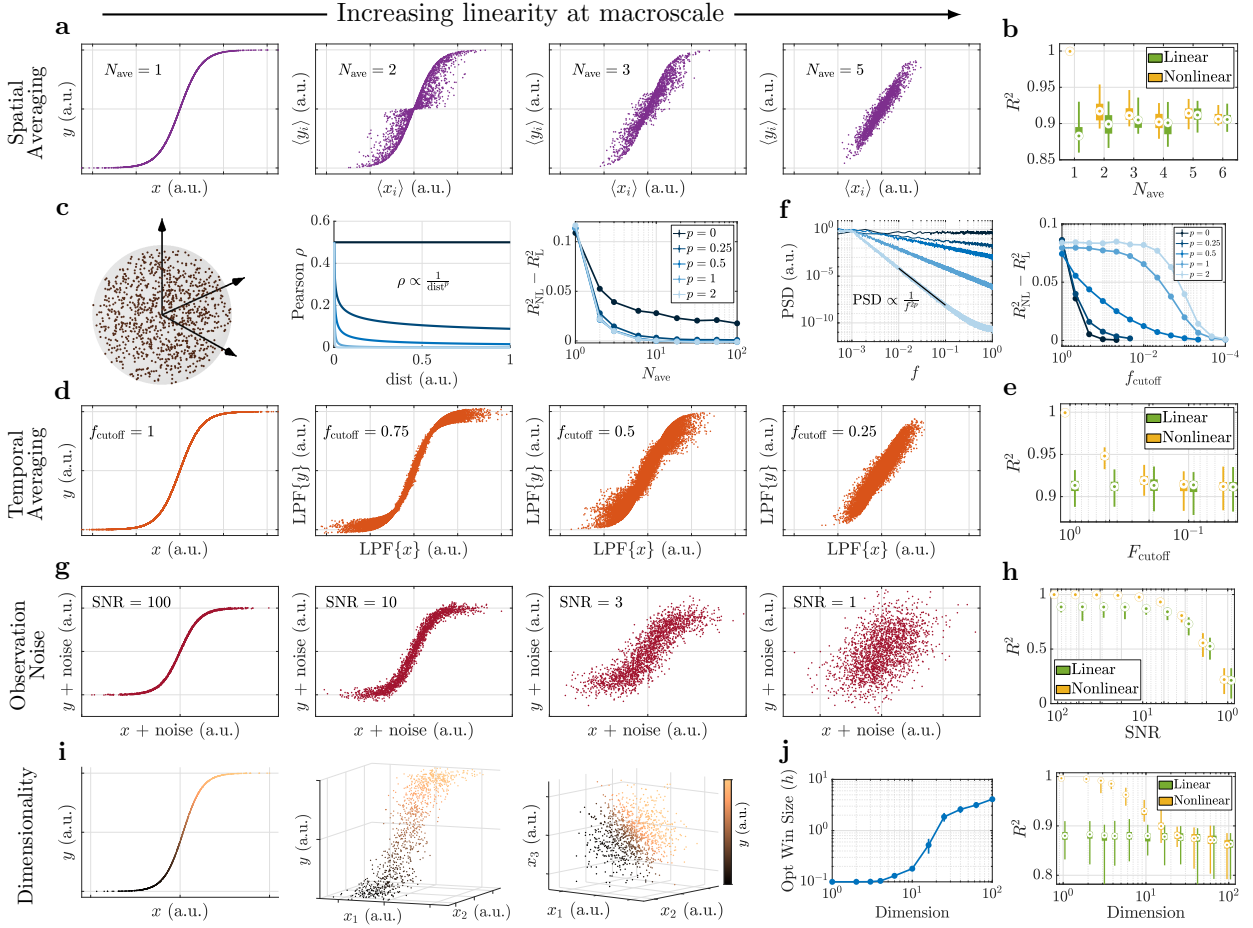

Figure 4: **The linearizing properties of macroscopic brain dynamics and of neuroimaging measurements.** (**a**) The effect of spatial averaging. For each panel, $N_{\text{ave}}$ pairs of signals $x_i(t), t = 1, \ldots, 2000$ were randomly and independently generated, $y_i(t) = \tanh(x_i(t))$ was calculated, and their averages $\langle x_i \rangle$ and $\langle y_i \rangle$ were computed. The quantities $\langle x_i \rangle$ and $\langle y_i \rangle$ possess a linear relationship as $N_{\text{ave}} \sim 5$ or higher. (**b**) The cross-validated $R^2$ of the optimal nonlinear (MMSE) and linear predictors for the $\langle x_i \rangle$-$\langle y_i \rangle$ relationships in panel (**a**). (**c**) The effect of spatial correlation on spatial averaging. Here, we assign $(x_i(t), y_i(t))$ pairs to spatial locations in a unit sphere (left) and make each $x_i(t)$ and $x_j(t)$ correlated in a manner that depends on their spatial distance (middle). The difference between nonlinear and linear $R^2$ always decays with $N_{\text{ave}}$ and vanishes if the correlation decays, even slowly, with distance (right). (**d**) The effect of temporal averaging. One pair of $x(t), y(t) = \tanh(x(t))$ is generated, independently over time, and passed through a low-pass Gaussian filter (LPF) with a cutoff frequency $f_{\text{cutoff}}$ that is normalized to the Nyquist frequency; thus, $f_{\text{cutoff}} = 1$ means no low-pass filtering. (**e**) Same as panel (**b**) but for the LPF$\{x\}$-LPF$\{y\}$ relationships in panel (**d**). (**f**) Similar to panel (**c**) but for temporal averaging. We varied the PSD decay rate of $x(t)$ (left) and then low-pass filtered $x(t)$ and $y(t) = \tanh(x(t))$ as in panel (**d**). The difference between the optimal linear and nonlinear $R^2$ eventually vanishes as $f_{\text{cutoff}}$ decreases, but it happens at smaller $f_{\text{cutoff}}$ for larger decay rates $p$. (**g**) The effect of observation noise. The quantities $x(t)$ and $y(t) = \tanh(x(t))$ are as in panel (**d**) and their additive noises are generated independently. (**h**) Same as panel (**e**) but for the $(x + \text{noise})$-$(y + \text{noise})$ relationships shown in panel (**g**). (**i**) The effect of dimensionality. The values $x_1(t), \ldots, x_n(t)$ are generated as in panel (**a**) but here $y(t) = \tanh(x_1(t) + \cdots + x_n(t))$ generates a one-dimensional nonlinearity in $n + 1$ dimensions. No noise is included; no spatial or temporal averaging is applied. (**j**) Right: similar to panels (**b, e, h**) except that a manifold-based (locally-linear) nonlinear predictor is used since the conditional density estimation required for MMSE loses accuracy in high dimensions with fixed number of data points (see Methods). Left: the optimal window size of the manifold-based predictor as a function of dimension $n$. As $n$ increases, the locally-linear predictor automatically chooses larger windows to be able to make reliable predictions, thereby effectively degrading to a globally linear predictor (see also Supplementary Figure 5). In all box plots, the center point, box limits, and whiskers represent the median, upper and lower quartiles, and the smallest and largest samples, respectively. Error bars in panels (**c, f, j**) represent one standard error of the mean.

In conclusion, the process of averaging over space, the process of averaging over time, the existence of observation noise, and the acquisition of limited data are each characteristic of macroscale brain dynamics or neuroimaging measurements, and can transform microscopically nonlinear dynamics into macroscopically linear ones. In reality, their effects are likely all combined, rendering the optimality of linear models in our comparisons not as unexpected as it might originally seem. This linearity has significant implications for computational neuroscience, as we discuss next.

## Discussion

**Summary.** In this work, we set out to test the hypothesis that macroscopic neural dynamics are nonlinear, and using linear models for them results in an inevitable loss of accuracy in exchange for simplicity. We thus compared linear and nonlinear models in terms of how well they can predict rsfMRI and rsiEEG data in a cross-validated prediction error (PE) system identification framework, where the quality of each model's fit was assessed by the variance and whiteness of its PE (residual). We found that linear models, and AR models in particular, achieve the lowest PE variance and highest PE whiteness, outperforming neural mass models, deep neural networks, manifold-based models, and the optimal MMSE predictors. Interestingly, the spatial (regional) distribution of the $R^2$ of the best model also shows significant differences across established cortical functional networks, a remarkably lower predictability of subcortical regions relative to cortical ones, and a close alignment between most methods (all but 'Linear w/ HRF') (Supplementary Figure 4). This distinction in predictability highlights significant differences in how spatio-temporally correlated the fMRI time series of different regions are, while the mechanistic physiological and technological reasons behind this distinction remains a warranted avenue for future research.

To further understand the possible causes of the optimality of linear models, we analyzed the effects of common elements of macroscopic neural dynamics: averaging over space and time, observation noise, and limited data samples. We showed that they can each counteract or mask the nonlinearities present at smaller scales. These linearizing effects will add up when combined, suggesting that linear models provide a useful choice for macroscopic neural dynamics at rest; of course, in certain experimental conditions, rigorous system identification methods might still uncover nonlinear dynamics in future studies.

The observed optimality of linear models for the resting state is accompanied by both challenges and opportunities. Having a linear model for neuroscience investigations is computationally ideal, given the extent to which the behavior of linear systems and their response to stimuli are mapped out. Nevertheless, to what extent these linearly-interacting macroscopic signals are informative of, and have a causal influence on, the underlying microscopic activity, remains unclear and represents an invaluable area for future investigation. Our observations also warrant the exploration and development of both linear and nonlinear models of macroscopic neural dynamics beyond those tested here and available in the literature.

**Connections to Prior Literature.** It is important to distinguish the pragmatic, modeling question that drove our analysis from the rather philosophical question of whether any signs of "nonlinearity" can be found in neuroimaging time series. The latter question has been extensively investigated [15–17,30], and often uses determinism or chaos as a proxy for nonlinearity. To answer our distinct modeling question, we used a system identification approach that allows for a direct, side-by-side comparison of linear and nonlinear models. In contrast, the aforementioned studies often resort to indirect, surrogate-based comparisons that rely on strong (and debated) assumptions about the constructed surrogates [31]. Also related to, but different from, our work are studies that seek to determine whether the end-to-end input-output mapping between stimuli and neuroimaging signals (EEG or BOLD) is nonlinear (e.g., see Refs. [32, 33]). Our focus here, however, is on the internal network dynamics of the brain, as well as on studies that have examined the performance of linear models *per se* in fitting neuroimaging time series (e.g., [34]) without a comparison to nonlinear models.

Our analysis of the linearizing effect of spatial correlation is also related to the large body of work investigating the effect of spatial correlations on the information content and decoding accuracy of neural population codes (see, e.g., [35]). As expected, the stronger the correlation between neurons, the weaker the linearizing effects of spatial averaging. However, nonlinearities can in principle have two opposing effects on the neural code. On the one hand, nonlinearities can substantially increase the computational complexity and expressivity of a neural network, making correlations beneficial for the neuronal encoding. On the other hand, if the expressivity is too high, the decodability of one neural population by another may decrease, potentially making the linearizing effects of low correlations favorable. Determining which effect dominates, and whether an optimal point exists at the levels of neural correlation observed, remain areas of future research *in vivo*.

**Results and Implications.** The implications of the linearity of brain dynamics are far-reaching. Linear systems fundamentally have a more limited repertoire of dynamic behaviors than nonlinear ones, excluding the possibility of multi-stability, chaos, limit cycles, or cross-frequency coupling, to name a few [36]. When driven by noise, linear systems act as linear filters that shape the power spectrum of their output (here, fMRI or iEEG time series) through their frequency response, essentially amplifying the frequency content near their resonance frequencies and dampening it elsewhere. Importantly, this effect of shaping the power spectrum of linear systems acts independently over different frequencies; in contrast, nonlinear systems can drive arbitrarily complex cross-frequency interactions [37].

The linearity of brain dynamics has even greater implications for network control [38, 39]. The design and analysis of optimal, robust, adaptive, and many other forms of control are much better understood in the context of linear systems than nonlinear ones. This contrast in tractability only grows for large-scale systems like the brain, thus motivating the recent surge of interest and advancements in using linear control theory in neuroscience [40–42]. Nonlinear models also present additional challenges beyond network control, including analytical and mechanistic understanding of their functionality, obtaining provable guarantees on their performance, and even hardware requirements for their use in chronic implantable devices. In this context, the present work shows that the favorable tractability and simplicity of linear models do not necessarily come at the often-presumed cost of model inaccuracy, and also provide the necessary tools for identifying the most accurate models for any datasets of interest.

In the analysis of fMRI data, we found that incorporating an HRF component in the model, instead of modeling the dynamics directly at the BOLD level, results in a loss of accuracy in linear models (see 'Linear (sparse)' vs. 'Linear w/ HRF'), and is almost ineffective in nonlinear models (see 'NMM' vs. 'NMM w/ HRF'). It was also in light of this observation that we did not include an HRF component in the majority of our models, such as the DNN or the manifold-based models. This lack of advantage of an explicit HRF component (within the specific context of modeling resting state fMRI dynamics using ODEs) is understandable on a number of grounds. First, in order to include an HRF component in the model, either one should learn the HRF from the data, such as in our 'linear w/ HRF' model, which will create marked model flexibility and therefore increase the likelihood of over-fitting; or, one should use a typical HRF, such as in our 'NMM w/ HRF' model, which is a source for additional error. Second, by including the HRF in the model, we ultimately seek to recover neural information that is lost through the HRF. This task is difficult, if not impossible, without a high signal-to-noise ratio as well as more accurate HRF models than those currently available. Finally, a linear autoregressive model can automatically capture a linear approximation of the HRF dynamics [43], precisely as present in the observed time series. Ultimately, our results encourage a side-by-side comparison of models with and without the inclusion of an HRF component in order to assess the costs and benefits of such inclusion for any datasets of interest.

A very similar argument also applies to including colored noise in the model, both for fMRI and iEEG. Even though the so-called "noise floor" in neural signals, over which neural oscillations are detected [44], has a clear power-law (1/f) PSD, its decay can be well modeled by white noise passing through a linear filter. This property of the 1/f noise is in fact why the AR models, which assume a white noise signal $\mathbf{e}(t)$, have prediction errors that are maximally white. This latter fact can be directly seen from Figures 2c and 3c, when noting that any model's prediction errors are, by construction, the model's estimate of the noise $\mathbf{e}(t)$ [21].

In addition to considering linear and nonlinear models, we wish to underscore the importance of the zero model. It is not uncommon in the modeling literature to assess the quality of a fitted model *per se*, without any grounds for comparison. For instance, our 'DNN (MLP)' model for fMRI had a median $R^2$ of about 14% and for some subjects it had a median $R^2$ (among all regions) of over 50%. Even more notably, the 'DNN (MLP)' model for iEEG had a median $R^2$ of over 97%. Without any comparisons, these numbers may suggest that these models are quite accurate; yet, as seen in Figures 2b and 3b, the predictive accuracy of these models are in fact lower than the zero model in fMRI, and indistinguishable from it in iEEG. The act of comparing to a baseline model is therefore an essential step in the assessment of any model's goodness-of-fit.

We restricted our analyses here in the main text to certain spatiotemporal resolutions for both fMRI (a coarse parcellation) and iEEG (a high sampling rate), naturally raising the question of how robust our findings are to our choices of resolutions. As shown in Supplementary Figures 11-16, our main finding (higher predictive power of linear autoregressive models over all other model families) holds across all resolutions tested. We do, however, observe certain differences between resolutions. In iEEG data, we observe that using lower sampling rates (and therefore longer time intervals) increases the benefit of modeling network interactions, even whilst lowering the $R^2$ values across all models. In fMRI data, we interestingly see that as we move towards more fine-grained parcellations and ultimately unparcelled data, (i) the simpler 'Linear (sparse)' model with less parameters gains advantage over the more populated 'VAR-3 (sparse)', and (ii) the overall $R^2$ values of all models is reduced, potentially due to the improvement in signal-to-noise ratio resulting from averaging in coarser parcellations.

**Methodological Considerations.** Despite the solid theoretical foundations of the PE method for system identification, our results may still beg a practical question: would the same system identification and side-by-side comparison procedure be able to identify nonlinear dynamics, should they actually exist in the time series? A direct answer to this question can be given, e.g., by applying the same procedure to simulated time series generated from a nonlinear model whose ground truth functional form we know. The result of such an analysis is provided directly in Supplementary Figures 2 and 3, but also indirectly in Figure 4. Note that in the latter, we compared the cross-validated predictive power of linear and (MMSE or manifold-based) nonlinear models in identifying the sigmoidal relationship $y = \sigma(x)$ and its variants, after averaging or noise addition. This relationship can be equally viewed as a nonlinear dynamical system $\dot{x} = \sigma(x)$, the nonlinearity of which was only identified until counteracted or masked by the four macroscopic effects we discussed therein.

As for the comparison of linear and nonlinear models, one might expect nonlinear models to perform at least as well as linear ones, but not worse, given that the space of nonlinear models includes all linear models as a special case. In our comparison, however, we saw that most of our nonlinear methods actually have a worse prediction performance than linear ones. This behavior can be understood in light of at least two facts. First, many nonlinear models, such as the neural mass models and DNNs, do not include linear models as a special case and have structural biases that can be a source of error if not consistent with the data [29]. Second, even nonlinear models that do not have structural biases and contain linear models as a special case, such as MMSE or manifold-based models, still have a marked flexibility relative to a linear model. These models indeed achieve the highest $R^2$ among nonlinear models both for fMRI and iEEG. However, their immense flexibility would in general lead to overfitting unless (i) abundant data is available, which is not currently the case for most neuroimaging modalities, or (ii) strict regularization is used, which is itself another source of bias and may not be consistent with the underlying data.

A further noteworthy aspect of our study specifically, and the prediction error framework more generally, is the focus on fitting the time series rather than its derivative statistics, such as the functional connectivity (FC) [45–47] or power spectral density [48]. While the choice of one approach over the other ultimately depends upon the anticipated use of the learned model, it is important to note that the mapping from dynamical systems to FC (or any other such statistic) is not a one-to-one mapping [49]. In fact, linear systems of the form $\mathbf{y}(t) - \mathbf{y}(t-1) = \mathbf{W}\mathbf{y}(t-1) + \mathbf{e}(t)$ with completely different $\mathbf{W}$ matrices can give rise to almost identical FC matrices (see Supplementary Figure 1). Therefore, when considering the accuracy of a general purpose model of the brain, the time series contains the maximum amount of information and thus provides the best target for model fitting.

One modeling approach that we did not employ in this study is dynamic causal modeling (DCM) [50]. The reason is that neither of the current variants of DCM are feasible, due to their computational complexity at the scale of our analysis: whole brain fMRI with $n = 116$ parcellations or large-scale iEEG with up to 175 and a median of 98 electrodes. The most efficient variant, spectral DCM for fMRI, for instance, is applicable to $\sim$ 30-40 nodes, whereas stochastic DCM (the most relevant to our study) is only applicable to much smaller systems. However, in light of our results thus far, the great computational complexity of the DCM approach, and thus its potential for overfitting, we would not expect its cross-validated $R^2$ to reach that of a linear model, although this comparison remains unknown at the present.

In this work, we demonstrated four properties of macroscopic neurodynamics that can counteract or mask microscopic nonlinearity. In doing so, we purposefully kept the discussion at a conceptual level and generally abstained from tying it to specific micro- or mesoscopic neural models, as doing so would require building on assumptions that our study explicitly seeks to avoid. For instance, it is currently unclear whether, and to what extent, the dynamics of the *mesoscopic* local field potentials or population firing rates that seem to be the main neural drivers of fMRI or iEEG are nonlinear and, if so, what the precise form of their nonlinearity is at each brain region. A warranted avenue for future research would be the re-analysis of the effects of spatial and temporal averaging, observation noise, and limited data samples on precise, data-driven models of mesoscopic brain dynamics, should they possess nonlinear interactions.

**Limitations.** Finally, we would like to highlight some of the limitations of the present study. First, it is important to note that the space of all nonlinear models in tens to hundreds of dimensions is intractably large, and the fact that our tested nonlinear models did not outperform linear ones is not a proof that no nonlinear model may ever do so. Our work thus seeks to provide rigorous evidence and methodology towards resolving the linear vs. nonlinear modeling dilemma in computational neuroscience, rather than a final resolution thereof. We can be confident, nevertheless, about the optimality of linear models at the pairwise level for fMRI or scalar AR level for iEEG given the equal or higher prediction power of linear regression relative to the optimal MMSE predictor. Moreover, our modeling framework is currently only applicable to resting state dynamics with no inputs, and has been tested on the two modalities of fMRI and iEEG. Inclusion of input signals for system identification of task fMRI/iEEG data requires accurate data-driven 'input models' of how experimental stimuli, as well as subjects' voluntary responses, influence the BOLD or LFP signals in each brain region, and is a highly warranted avenue for future research [51, 52]. Under intensive task conditions, moreover, it is more likely, or perhaps certain, to observe nonlinearities at least in the form of saturation effects in the BOLD/LFP signal. However, the precise form and extent of this nonlinearity needs to be determined using rigorous system identification routines.

In conclusion, our work sought to ask the often unasked question of whether the brain is macroscopically linear. We compared various linear and nonlinear families of models to determine their relative advantages and disadvantages. Our findings show that simple linear models explain the rsfMRI and rsiEEG data as well as, or even better than, an array of nonlinear ones, thus challenging the commonly held, yet untested assumption of higher accuracy of nonlinear models. However, the costs and benefits of nonlinear models are ultimately case-specific. Therefore, instead of offering a universal recommendation on the preferable choices for the modeling of neural dynamics, we rather provide the groundwork for rigorous in-vestigation and informed decision-making in the context of rsfMRI/rsiEEG. When feasible, following a similar system identification routine is always recommended for computational modeling of any datasets of interest, in order to ensure the optimal fit of the models used for subsequent analysis or design.

## Methods

**Data and pre-processing.** For the fMRI analysis, we used ICA-FIX resting state data from the S1200 Human Connectome Project release [53, 54]. rsfMRI images were collected with the following parameters: TR = 720 ms, TE = 33.1 ms, flip angle = 52 deg, FOV = 208x108 mm, matrix = 104x90, slice thickness = 2.0 mm, number of slices = 72 (2.0 mm isotropic), multi factor band = 8, and echo spacing = 0.58 ms. Brains were normalized to fslr32k via the MSM-AII registration and global signal was removed. No bandpass filtering was performed (see Supplementary Note 1). Finally, we removed subjects from further analysis if any of their four resting state scans had excessively large head motion, defined by having frames with greater than 0.2 mm frame-wise displacement or a derivative root mean square (DVARS) above 75. Also, subjects listed in [55] under "3T Functional Preprocessing Error of all 3T RL fMRI runs in 25 Subjects" or "Subjects without Field Maps for Structural scans" were removed, leaving a total of 700 subjects that were used for all the analyses. We parcellated the brain into 100 cortical regions (Schaefer 100x7 atlas [56]) and 16 subcortical ones (Melbourne Scale I atlas [57]).

For iEEG preprocessing, raw data from the RAM data set we have published on previously [58–60] was segmented into task free epochs from either before or after task completion that were at least 5 minutes in length. This process resulted in a total of 283 recordings from 122 subjects. Data were then downsampled to the lowest sampling rate used across recording sites (500 Hz). Electric line noise and its harmonics at 60, 120, and 180 Hz were filtered out using a zero phase distortion 4-th order stop-band Butterworth filter with a 1 Hz width. This procedure was implemented using the *butter()* and *filtfilt()* functions in MATLAB. We then rejected noisy channels that were either (i) marked as noisy in the RAM dataset notes, (ii) had a line length greater then three times the mean, (iii) had $z$-scored kurtosis greater than 1.5, or (iv) had a $z$-scored power-spectral density dissimilarity measure greater than 1.5. The dissimilarity measure used was the average of one minus the Spearman's rank correlation with all channels. Data were then demeaned and detrended. Channels were grouped according to whether they were grid or depth electrodes, and then common average referenced within each group. Following the common average referencing step, plots of raw data and power spectral densities were visually inspected by an expert researcher with 6 years of experience working with electrocorticography data to ensure that data were relatively clean.

**Computing.** All the computations whose run time was measured and reported in Figures 2e and 3e were performed on the CUBIC cluster at the University of Pennsylvania, using 1 CPU core and 16 or 64 GB of memory per fMRI or iEEG computing jobs, respectively.

**Linear and nonlinear families of models.** The continuous-time dynamics in Eq. (1) is first discretized. With a slight abuse of notation, we also represent the discretized dynamics as

$$\mathbf{x}(t) - \mathbf{x}(t-1) = f(\mathbf{x}(t-1)) + \mathbf{e}_1(t), \quad \mathbf{x}(0) = \mathbf{x}_0 \quad (4a)$$
$$t = 1, \ldots, N$$
$$\mathbf{y}(t) = h(\mathbf{x}(t)) + \mathbf{e}_2(t), \qquad t = 0, \ldots, N \quad (4b)$$

where the time index $t$ is now an integer, for simplicity of notation, but the discretization step size is always equal to 0.72 seconds for the HCP data (equivalent to 1 TR) and 2 milliseconds for the RAM data. This choice means that, e.g., the map $f$ in Eq. (4a) equals 1 time step multiplied by the map $f$ in Eq. (1a), and $T = 864s$ in Eq. (1a) corresponds to $N = 1200$ in Eq. (4a) for the fMRI data. Recall that in this general form, the noise signals $\mathbf{e}_1(t)$ and $\mathbf{e}_2(t)$ can have arbitrary statistics, including white or colored PSD. We then learn the dynamics in Eq. (4) using the following families of models. The hyper-parameters used for each model are listed in Table 1 (see 'Hyper-parameter selection' below and Supplementary Figures 9 and 10 for details).

*Linear models with states at the BOLD/LFP level ('Linear (dense)', 'Linear (sparse)'):* This model is our simplest. In it, we let $\mathbf{y}(t) = \mathbf{x}(t)$, modeling the dynamics directly at the BOLD/LFP level. This also allows for combining the noise signals $\mathbf{e}_1(t)$ and $\mathbf{e}_2(t)$ into a single noise signal $\mathbf{e}(t)$, which is then taken to be white. These simplify Eq. (4) to

$$\mathbf{y}(t) - \mathbf{y}(t-1) = f(\mathbf{y}(t-1)) + \mathbf{e}(t). \quad (11)$$

If we further let $f(\mathbf{y}(t)) = \mathbf{W}\mathbf{y}(t)$ be linear, then we get Eq. (5) (see Table 1) where $\mathbf{W}$ is an $n$-by-$n$ matrix of effective connectivity between brain regions. We fit and compare this model both when $\mathbf{W}$ is dense and when it is sparse. The latter is motivated by the facts that (i) from a mechanistic perspective, an important property of brain networks and other large-scale complex networks is their sparsity; while (ii) from a machine learning perspective, regularization and reducing the number of free parameters in a model can prevent over-fitting and improve generalization. To promote sparsity, we use standard 1-norm (LASSO) regularization with a $\lambda$ hyper-parameter that is tuned separately for fMRI and iEEG.

**Table 1:** Linear and nonlinear families of models used in this study. The marks † and ‡ indicate, respectively, that a method is used only for fMRI or iEEG. See Methods for a description of each model.

| Label | Title | Equation | Hyper-parameters |
|---|---|---|---|
| Linear (dense) | Linear models with states at the BOLD/LFP level | $\mathbf{y}(t) - \mathbf{y}(t-1) = \mathbf{W}\mathbf{y}(t-1) + \mathbf{e}(t)$ (5) | None |
| Linear (sparse) | | | $\lambda = 0.95$ (fMRI) $\lambda = 1.2$ (iEEG) |
| Linear (pairwise)† | | $y_i(t) - y_i(t-1) = w_{ij} y_j(t-1) + e_i(t), i, j = 1, \ldots, n$ | None |
| AR-2 (sparse)† | Linear autoregressive models | $\mathbf{y}(t) - \mathbf{y}(t-1) = \mathbf{W}\mathbf{y}(t-1) + \mathbf{D}_2 \mathbf{y}(t-2)$ $+ \mathbf{D}_3 \mathbf{y}(t-3) + \cdots$ $+ \mathbf{D}_d \mathbf{y}(t-d) + \mathbf{e}(t)$ (6) | $d=2, \lambda=0.95$, diagonal $\mathbf{D}_2$ |
| VAR-2 (sparse)† | | | $d=2, \lambda=0.9$ |
| AR-3 (sparse)† | | | $d=3, \lambda=0.5$, diagonal $\mathbf{D}_2, \mathbf{D}_3$ |
| VAR-3 (sparse)† | | | $d=3, \lambda=0.35$ |
| AR-100 (sparse)‡ | | | $d=100, \lambda=1.5$ |
| AR-100 (scalar)‡ | | | $d=102$ |
| Linear w/ HRF† | Linear models with states at the neural level | $\mathbf{x}(t) - \mathbf{x}(t-1) = \mathbf{W}\mathbf{x}(t-1) + \mathcal{G}_1(q)\hat{\mathbf{e}}_1(t)$ (7a) $\mathbf{y}(t) = \mathcal{H}(q)\mathbf{x}(t) + \mathcal{G}_2(q)\hat{\mathbf{e}}_2(t)$ $\mathcal{H}(q) = \sum_{p=1}^{n_h} \mathrm{diag}(\mathbf{H}_{:,p})q^{-p}$ (7b) $\mathcal{F}_1(q) = \mathbf{I} - \mathcal{G}_1^{-1}(q) = \sum_{p=1}^{n_\phi} \mathrm{diag}(\mathbf{\Phi}_{:,p})q^{-p}$ (7c) $\mathcal{F}_2(q) = \mathbf{I} - \mathcal{G}_2^{-1}(q) = \sum_{p=1}^{n_\psi} \mathrm{diag}(\mathbf{\Psi}_{:,p})q^{-p}$ (7d) | $n_h = n_\phi = n_\psi = 5, \lambda = 11$ |
| Subspace | Linear models with abstract data-driven states | $\mathbf{x}(t) - \mathbf{x}(t-1) = \mathbf{W}\mathbf{x}(t-1) + \mathbf{e}_1(t)$ $\mathbf{y}(t) = \mathbf{C}\mathbf{x}(t) + \mathbf{e}_2(t)$ $\mathrm{Cov}\left(\begin{bmatrix}\mathbf{e}_1(t)\\ \mathbf{e}_2(t)\end{bmatrix}\right) = \begin{bmatrix}\mathbf{Q} & \mathbf{M}\\ \mathbf{M}^T & \mathbf{R}\end{bmatrix}$ (8) | $s=1, r=3, n=25$ (fMRI) $s=11, r=49, n=436$ (iEEG) |
| NMM | Nonlinear neural mass models | $\mathbf{y}(t) - \mathbf{y}(t-1) = (\mathbf{W}\psi_\alpha(\mathbf{y}(t-1)) - \mathbf{D}\mathbf{y}(t-1))\Delta_T + \mathbf{e}(t)$ | MINDy default (fMRI) $\lambda_1 = \lambda_2 = 0.2, \lambda_3 = 2,$ $\lambda_4 = 0.5$ (iEEG) |
| NMM w/ HRF† | | $\mathbf{x}(t) - \mathbf{x}(t-1) = (\mathbf{W}\psi_\alpha(\mathbf{x}(t-1)) - \mathbf{D}\mathbf{x}(t-1))\Delta_T$ $+ \mathbf{e}_1(t)$ $\mathbf{y}(t) = \mathcal{H}(q)\mathbf{x}(t) + \mathbf{e}_2(t)$ (9) | MINDy default |
| DNN (MLP) | Nonlinear models via multi-layer perceptron deep neural networks | $\mathbf{y}(t) - \mathbf{y}(t-1) = f(\mathbf{y}(t-1), \ldots, \mathbf{y}(t-d)) + \mathbf{e}(t)$ (10) | $d=1, D=6, W=2$ (fMRI) $d=6, D=4, W=26$ (iEEG) |
| DNN (CNN) | Nonlinear models via convolutional deep neural networks | $\mathbf{y}(t) - \mathbf{y}(t-1) = f(\mathbf{y}(t-1), \ldots, \mathbf{y}(t-d)) + \mathbf{e}(t)$ | $d=17, D=2, l_{\mathrm{filt}}=7,$ $n_{\mathrm{filt}}=11, n_{\mathrm{pool}}=4,$ $p_{\mathrm{drop}}=0.4$ (fMRI) $d=11, D=7, l_{\mathrm{filt}}=2,$ $n_{\mathrm{filt}}=13, n_{\mathrm{pool}}=1,$ $p_{\mathrm{drop}}=0.5$ (iEEG) |
| LSTM (IIR) | Nonlinear models via long short-term memory recurrent neural networks | $\mathbf{y}(t) - \mathbf{y}(t-1) = f(\mathbf{y}(t-1), \ldots, \mathbf{y}(0)) + \mathbf{e}(t)$ | $W=12$ (fMRI) $W=1$ (iEEG) |
| LSTM (FIR) | | $\mathbf{y}(t) - \mathbf{y}(t-1) = f(\mathbf{y}(t-1), \ldots, \mathbf{y}(t-d)) + \mathbf{e}(t)$ | $d=1, W=16$ (fMRI) $d=7, W=1$ (iEEG) |
| Manifold | Nonlinear manifold-based models | $\mathbf{y}(t) - \mathbf{y}(t-1) = f(\mathbf{y}(t-1), \ldots, \mathbf{y}(t-d)) + \mathbf{e}(t)$ | $d=1, h=830$ (fMRI) $d=7, h=1.2 \times 10^4$ (iEEG) |
| MMSE (pairwise)† | Nonlinear minimum mean squared error models (optimal) | $y_i(t) - y_i(t-1) = \mathrm{E}[y_i(t) - y_i(t-1) \mid y_j(t-1)], i, j = 1, \ldots, n$ | $N=280, \beta=0.156$ |
| MMSE (scalar)‡ | | $y_i(t) - y_i(t-1) = \mathrm{E}[y_i(t) - y_i(t-1) \mid y_i(t-1), \ldots,$ $y_i(t-d)], i = 1, \ldots, n$ | $d=15, N=300, \beta=0.007$ |
| Zero | Zero model | $\mathbf{y}(t) - \mathbf{y}(t-1) = \mathbf{e}(t)$ | None |

*Linear autoregressive models ('AR-2 (sparse)', 'VAR-2 (sparse)', 'AR-3 (sparse)', 'VAR-3 (sparse)', 'AR-100 (sparse)', 'AR-100 (scalar)'):* motivated by the long history of AR models in neuroscience [23, 61, 62], here we extend Eq. (5) to Eq. (6) (see Table 1) for an 'AR-$d$' model. The number of lags $d$ was tuned separately for fMRI and iEEG, and the matrix $\mathbf{W}$ is either made sparse using LASSO or enforced to be diagonal. Note that the latter results in $n$ scalar AR models at each node which are completely decoupled from each other. We restricted the matrices $\mathbf{D}_2, \mathbf{D}_3, \ldots$ to be diagonal in 'AR' models but not so in full vector auto-regressive ('VAR') models. In both cases, we use LASSO regularization to promote sparsity in the regressors, signified by the '(sparse)' suffix in method identifiers, with the regularization hyper-parameter $\lambda$ chosen optimally and separately for each model (cf. 'Hyper-parameter selection' below). In general, we found that $\lambda$ is a moderately sensitive parameter, more so for the whiteness of residuals than $R^2$ (cf. Supplementary Figure 7 for an example).

*Linear models with states at the neural level ('Linear w/ HRF', only applicable to fMRI data):* A standard step in the computational modeling of fMRI dynamics is to incorporate a model of the hemodynamic response function (HRF), and to separate the underlying neuronal variables from the observed BOLD signals. In this family of models, we thus separate the states $\mathbf{x}$ from the outputs $\mathbf{y}$, while keeping a one-to-one relationship between the two ($m = n$). We then let the latter be a filtered version of the former through the HRF. For generality and given the natural and important variability of HRF across the brain [63, 64], we allow the HRF to vary regionally and learn it from the data for all regions in addition to the effective connectivity matrix $\mathbf{W}$. Furthermore, for the sake of generality, we allow both $\mathbf{e}_1(t)$ and $\mathbf{e}_2(t)$ to be colored, with power spectral densities that can also be different between regions and are learned from data. Note that this choice includes, as a special case, white $\mathbf{e}_1(t)$ and $\mathbf{e}_2(t)$. The result is a highly flexible linear model given by Eq. (7) (see Table 1). Since LASSO regression produced the best results in our BOLD-level linear models, we use LASSO to promote sparsity in $\mathbf{W}$ here. $\mathcal{H}(q)$ is a diagonal matrix whose $(i, i)$ entry is a linear finite-impulse response (FIR) approximation of the HRF in region $i$, parameterized as in Eq. (7b) ($q^{-1}$ is the standard delay operator, such that $q^{-1}x(t) = x(t-1)$, see [21]). Similarly, $\mathcal{G}_1$ and $\mathcal{G}_2$ are diagonal filters, parameterized by the inverse FIR forms in Eq. (7c)-(7d). Since the state vector $\mathbf{x}(t)$ is not measured, we learn this model by iterating between state estimation and parameter estimation in an expectation-maximization (EM)-like manner. Note that the presence of filters increases the effective state dimension of the system to $n \cdot \max\{n_\phi + 1, n_h + n_\psi\}$, significantly increasing the computational complexity of the state estimation step. The final model is taken from the EM iteration with the highest (training) $R^2$.

*Linear models with abstract data-driven states ('Subspace'):* The previous model, despite and because of its extreme generality and flexibility, has a very large state dimension and is extremely difficult to fit. If we forgo the physiological interpretability of the states, then significantly simpler and lower-dimensional models of the form in Eq. (8) (see Table 1) can be learned via subspace identification methods [21]. Unlike the model above, states represent abstract low-dimensional regularities within the data, with a dimension $m$ that is chosen optimally for each data type. The noise sequences $\mathbf{e}_1(t)$ and $\mathbf{e}_2(t)$ are assumed to be white but can be correlated, and the covariance matrices $\mathbf{Q}$, $\mathbf{M}$, and $\mathbf{R}$ are also learned from data. We note here that the subspace method used for learning this model is not technically a PE method, but we still use the PE framework for the cross-validated computation of $R^2$ and whiteness of residuals.

*Nonlinear neural mass models ('NMM', 'NMM w/ HRF'):* Learning of the models above, except for the 'Linear w/ HRF', involves a convex optimization that can be efficiently solved to find its unique global optimum. In contrast, the learning of nonlinear models is less straightforward. Recently, Singh *et al.* [65] developed an algorithm called MINDy that uses state-of-the-art optimization techniques for learning a neural mass model of the form in Eq. (9) (see Table 1) using rsfMRI data. In this model, $\mathbf{x}(t)$ has the same dimension as $\mathbf{y}(t)$ (one neural mass per brain region), $\Delta_T$ is the sampling time, $\mathbf{W}$ is a sparse connectivity matrix, $\mathbf{D}$ is a diagonal self decay matrix, $\psi_{\boldsymbol{\alpha}}(\cdot)$ is an element-wise sigmoidal nonlinearity whose steepness is determined by each element of the vector $\boldsymbol{\alpha}$ (which is also the same size as $\mathbf{x}$), and $\mathcal{H}(q)$ is a scalar linear HRF that is the same and fixed *a priori* for all regions. The associated toolbox that we use allows the user to either deconvolve $\mathbf{y}(t)$ using a canonical HRF to obtain the state $\mathbf{x}(t)$ ('NMM w/ HRF'), or set $\mathcal{H}(q) = 1$ and directly fit the model to $\mathbf{y}(t)$ ('NMM'). We use both methods for fMRI data but only the latter for iEEG. Since the MINDy algorithm was originally tuned for fMRI, we re-tune its regularization hyper-parameters $\lambda_1, \ldots, \lambda_4$ for use with iEEG data.

*Nonlinear models via multi-layer perceptron deep neural networks ('DNN (MLP)'):* Here we use a model of the form in Eq. (11) for fMRI and train a rectified linear unit (ReLU) MLP DNN to approximate the function $f(\cdot)$. The structure of the DNN consists of an input layer, $D$ ReLU layers, each preceded with fully connected and batch normalization layers and succeeded with a 50% dropout layer, a final fully connected layer, and the output layer. Given the importance of AR lags in the modeling of iEEG, for this modality we generalize Eq. (11) as Eq. (10) (see Table 1 for the latter) and similarly approximate $f(\cdot)$ using an MLP DNN. We use MATLAB's Deep Learning Toolbox for the training and evaluation of the DNN and tune the depth $D$ and width $W$ of the DNN separately for fMRI and iEEG (see 'Hyper-parameter selection' below).

*Nonlinear models via convolutional deep neural networks ('DNN (CNN)'):* Given the recent success of CNNs in complex learning problems, we also included a model similar to 'DNN (MLP)' but with a CNN to approximate the function $f(\cdot)$. The network consists of an input layer, $D$ one-dimensional convolutional layers (convolving over time using $n_{\text{filt}}$ filters of size $l_{\text{filt}}$) each succeeded by a batch normalization layer, a ReLU layer, and an average pooling layer with a pool size of $n_{\text{pool}}$, a final dropout layer with probability $p_{\text{drop}}$, a fully connected layer, and the output layer. Spatial convolution was not included in the model, as is the standard in modeling dynamical systems with CNNs, due to the arbitrary nature of channel numbering. Temporal convolution is nevertheless the basis of this model and we thus consider $d > 1$ autoregressive lags for both fMRI and iEEG.

*Nonlinear models via long short-term memory neural networks ('LSTM (IIR)', 'LSTM (FIR)'):* The above DNN models are inherently static (i.e., feedforward), whereas various recurrent neural network architectures have also been proposed for directly modeling dynamical systems. One of the most successful such architectures are LSTMs which we implemented here in two forms: infinite impulse response (IIR) and finite impulse response (FIR). These two forms correspond, respectively, to the two common sequence-to-sequence and sequence-to-one forms of modeling time series using LSTMs. In both cases, the network consists of an input layer, a layer of $W$ LSTM units, a fully connected layer, and an output layer. The difference is that in the IIR model, the network is initialized once at time 0 and run forward, continuously receiving $\mathbf{y}(t-1)$ as input and generating $\mathbf{y}(t) - \mathbf{y}(t-1)$ as output. Each output, therefore, depends on the entire history of the inputs. In the FIR model, on the other hand, the model is initialized and run forward once for each time point $t$, receiving only $\mathbf{y}(t-d), \ldots, \mathbf{y}(t-1)$ as input when predicting $\mathbf{y}(t) - \mathbf{y}(t-1)$.

*Nonlinear manifold-based models ('Manifold'):* Consider Eq. (11) or Eq. (10) and assume, for simplicity, that $f$ is differentiable. Each of these systems of equations consists of $n$ scalar equations, each of which defines a manifold (surface) in $n+1$-dimensional space. Various methods have been developed in the machine learning and system identification literature [66, 67] and used in computational neuroscience [18, 68] to capitalize on the fact that in the small vicinity of a point, the manifold can be approximated by a linear hyperplane tangential to it at that point. Here, we use the simple method of local polynomial modeling of order 1 [66] where, in the vicinity of any test point $\mathbf{y}_0$, we approximate

$$f(\mathbf{z}) \simeq \mathbf{c} + \mathbf{W}(\mathbf{z} - \mathbf{z}_0)$$

and fit $\mathbf{c}$ and $\mathbf{W}$ using the training points that are "close" to $\mathbf{y}_0$. Same applies for approximating $f$ in Eq. (10). To define which points are close, this method uses a window function with a width hyper-parameter $h > 0$ that determines how wide that window should be. We tuned $h$ separately for fMRI and iEEG (see 'Hyper-parameter selection' below), giving rise to values that are so large that they essentially result in a globally linear model (Supplementary Figure 5). The value of $h$ was independently optimized for the computations reported in Figure 4i-j, as described below.

*Nonlinear minimum mean squared error models (optimal) ('MMSE (pairwise)', 'MMSE (scalar)'):* The models in Eq. (11) (for fMRI) or Eq. (10) (for iEEG) ultimately define a stochastic mapping from $\mathbf{y}(t-1)$ or $(\mathbf{y}(t-1), \ldots, \mathbf{y}(t-d))$ to $\mathbf{y}(t) - \mathbf{y}(t-1)$ such that observing the values of the former provides information to predict the latter. It is not hard to show that for two random variables $U$ and $V$, the optimal (i.e., minimum variance) prediction of $U$ given $V = v$ is given by its conditional expectation $\hat{u} = \mathrm{E}[U|V = v]$ known as the MMSE prediction [69]. Therefore, the optimal prediction of $\mathbf{y}(t) - \mathbf{y}(t-1)$ given $\mathbf{y}(t-1)$ or $(\mathbf{y}(t-1), \ldots, \mathbf{y}(t-d))$ is given by $\mathrm{E}[\mathbf{y}(t) - \mathbf{y}(t-1)|\mathbf{y}(t-1)]$ and $\mathrm{E}[\mathbf{y}(t) - \mathbf{y}(t-1)|\mathbf{y}(t-1), \ldots, \mathbf{y}(t-d)]$, respectively. Due to its optimality, it provides a theoretical upper bound on the achievable accuracy of *any* nonlinear model. The difficulty in calculating this estimate, however, is the estimation of the conditional distribution of $\mathbf{y}(t) - \mathbf{y}(t-1)$ given an observation of $\mathbf{y}(t-1)$ or $(\mathbf{y}(t-1), \ldots, \mathbf{y}(t-d))$. Without imposing additional assumptions (e.g., Gaussianity), this task is not feasible in $n \sim 100$ dimensions with our limited data points per recording segment. However, this distribution is indeed feasible (i) on a pairwise basis, giving us the optimal predictions $\mathrm{E}[y_i(t) - y_i(t-1)|y_j(t-1)]$ for all pairs $i, j = 1, \ldots, n$, or (ii) on a scalar AR basis, yielding the optimal predictions $\mathrm{E}[y_i(t) - y_i(t-1)|y_i(t-1), \ldots, y_i(t-d)]$ separately for each $i = 1, \ldots, n$. We use the former for fMRI and the latter for iEEG. To estimate this conditional distribution for fMRI, we use a Gaussian window with a standard deviation equal to $\beta$ times the range of $y_j(t)$ in the training data to detect the training points close to each test $y_j(t-1)$ and then use an $N$-point weighted histogram to estimate the (conditional) distribution. The case for iEEG is similar. $\beta$ and $N$ are hyper-parameters that are tuned separately for fMRI and iEEG (see 'Hyper-parameter selection' below).

*Zero model ('Zero'):* So far, we have discussed several families of models. Comparisons among them will provide a clear picture of which family provides the best fit to the data *relative to* the others. Note though that this process does not necessarily imply that the best model is good in any absolute sense. In other words, all models may be estimating $\hat{\mathbf{y}}(t|t-1)$ at chance level or lower. Therefore, we also consider the zero model (a.k.a. zero-order hold, naive, or random walk)

$$\mathbf{y}(t) - \mathbf{y}(t-1) = \mathbf{e}(t)$$

with the trivial estimate $\hat{\mathbf{y}}(t|t-1) = \mathbf{y}(t-1)$. Note that this expression corresponds to Eq. (11) with $f(\mathbf{y}(t-1)) = \mathbf{0}$ and is only meant to provide a baseline for comparison, not to act as a formal model itself. Also note that this model is different from, and often performs better than, the constant predictor $\hat{\mathbf{y}}(t|t-1) = \bar{\mathbf{y}}$ which constitutes the denominator of $R^2$.

**Hyper-parameter selection.** For all models that involve the choice of a design hyper-parameter, we simultaneously optimized over all the hyper-parameters using stochastic gradient descent (SGD) with minibatch, separately for fMRI and iEEG. Let $N_{\text{param}}$ denote the number of hyper-parameters in any of the models. Starting from an initial estimate of the hyper-parameter vector, in each iteration, $3^{N_{\text{param}}}$ hyper-parameter vectors were generated, constituting a hyper-cubic mesh around the current hyper-parameter estimate. For integer-valued hyper-parameters, we moved 1 point in each direction while for real-valued hyper-parameters, we moved $10^{-6}$ units. Using a minibatch of randomly selected data segments, the mean-over-minibatch of the median-over-regions of the model $R^2$ was computed and maximized over the mesh. The random minibatch selection was independent between mesh points and between iterations. For integer-valued hyper-parameters, their value was updated to that of the maximizing mesh point. For real-value hyper-parameters, a gradient-ascent step was taken in the direction of the largest $R^2$. The process was repeated until the hyper-parameters stopped having a consistent decrease/increase and hovered around a steady state value (which always happens due to the stochastic nature of SGD) and/or the $R^2$ stopped having a consistent increase. The hyper-parameter and $R^2$ values throughout the process are shown in Supplementary Figures 9 and 10 for fMRI and iEEG data, respectively, and the final values of the hyper-parameters selected for each model are reported in Table 1. Note that the initial hyper-parameter estimates were chosen based on prior experience, not randomly, which is why they are often very close to or the same as the final values.

**Nonlinear predictors used for the analysis of the linearizing effects of macroscopic dynamics.** Our discussions of linear and nonlinear models and their hyper-parameters so far applies to the comparisons shown in Figures 2 and 3 on neuroimaging time series. In our numerical analysis of the linearizing effects of macroscopic brain dynamics in Figure 4, we also construct linear and nonlinear predictors and compute their $R^2$. The linear predictor is always a simple linear regression model, while the nonlinear predictor is the MMSE predictor for two-dimensional predictions (Figure 4a-h), and the manifold-based predictor for higher-dimensional predictions (Figure 4i-j). The MMSE predictor was as above, except that $\beta$ was adjusted as $0.02 + 0.02/\text{SNR}$ for Figure 4g-h. For the manifold-based predictor, we used a Gaussian window and swept logarithmically over its hyper-parameter $h$ from 0.1 to 10 in every iteration and chose the value of $h$ that gave the largest $R^2$. Figure 4j-left panel shows the average of the resulting optimal $h$ for 100 iterations.

**Cross-validation.** For the comparisons of Figure 2 on HCP data, we performed the cross-validation as follows, with slightly different procedures for brain-wide and regional methods. For the brain-wide methods, for each of the 700 subjects, we split each of the 4 resting scans of that subject into 2 halves, giving a total of 8 segments, each of length 600 samples. All of our methods were then applied using an 8-fold cross-validation where each time one of the 8 segments was used for testing and the remaining 7 were used for training. For pairwise methods, we were forced to lower the sample size due to the extremely high computational complexity of the MMSE predictor. Therefore, instead of each of the above 8 segments (per subject), we used the second quarter of that segment, giving us still an 8-fold cross-validation but on segments of length 150 samples each.

For the comparisons of Figure 3 on RAM data, we first split each of the 283, 5-minute recordings into 8490, 10-second segments. Even though having longer segments would in principle benefit model fitting, 10-second segments ensured that all of our methods could run using the 64 GB available memory per node on the CUBIC cluster. From the 8490 segments, those that contained any NaN entries (316 segments) or for which the subspace method produced NaN predictions (30 segments, due to the bad conditioning of the $\boldsymbol{\Phi}$ matrix therein) were removed from further analysis. Since each recording is already split into 30 segments, and due to the large number of segments, we performed only a single-fold cross-validation on each segment, with the first 8 seconds used for training and the final 2 second used for cross-validation.

**Multivariate test of whiteness.** A standard measure of the goodness of fit in the prediction error method is the whiteness of residuals, measuring the extent to which all temporal structure (i.e., dynamics) in the data has been captured by the model. Note that a multivariate time series $\mathbf{e}(t)$ is "white" if it has no statistical dependence across time (i.e., $\mathbf{e}(s)$ and $\mathbf{e}(t)$ are independent if $s \neq t$) even though it can have arbitrary statistical dependence across channels (i.e., $e_i(t)$ and $e_j(t)$ can be dependent at the same time $t$). Parametric ($\chi^2$) statistical tests have been devised for multivariate whiteness, such as the classical Box-Pierce portmanteau test [70] and its modifications by Ljung & Box [71] and Li & McLeod [72]. Under strong assumptions, all of these tests have a statistic $Q$ (defined slightly differently between them) that is asymptotically (at infinite samples) $\chi^2$ distributed. In our datasets we found, however, that $Q$ is not $\chi^2$ distributed, and therefore we use randomization to generate the true null distribution of $Q$ by shuffling the time indices of $\mathbf{e}(t)$ 100 times, computing $Q$ for each of them, and computing the 95th percentile of the randomized $Q$ values as the threshold $Q_{\text{thr}}$ for significance. We use the original definition of $Q$ [70],

$$Q = (N - M) \sum_{i=1}^{M} \text{trace}\left(\hat{\mathbf{R}}_{\mathbf{e}}(i)^T \hat{\mathbf{R}}_{\mathbf{e}}(0)^{-1} \hat{\mathbf{R}}_{\mathbf{e}}(i) \hat{\mathbf{R}}_{\mathbf{e}}(0)^{-1}\right)$$

where $N$ is the number of (test) samples, $M$ is the number of cross-correlation lags, and

$$\hat{\mathbf{R}}_e(i) = \frac{1}{N-M} \sum_{t=0}^{N-M-1} \mathbf{e}(t+i)\mathbf{e}(t)^T, \quad i = 0, 1, \ldots, M.$$

is a finite-sample estimate of the cross-correlation matrix between channels of $\mathbf{e}(t)$ at lag $i$. Since in practice $\hat{\mathbf{R}}_{\mathbf{e}}(0)$ may be singular or near-singular, we use the pseudo-inverse of $\hat{\mathbf{R}}_{\mathbf{e}}(0)$ instead of its inverse in computing $Q$. Finally, only in the case of "pairwise" fMRI models where the residuals are inherently univariate, we use the simpler $\chi^2$ test of whiteness for univariate time series [21, §16.6].

**Estimation of rsfMRI SNR.** Here we describe our method for the estimation of rsfMRI time series scanner noise and the resulting SNR reported in Supplementary Figure 8. From the 700 subjects used for the study, 50 were selected uniformly at random, and for each selected subject, one of their 4 rest scans was selected also uniformly at random. The following was then performed for each of the 50 subject-scans. The rest scan was motion corrected using intra-modal linear registration with 6 degrees of freedom (in general, we kept the amount of pre-processing as minimal as possible throughout the SNR estimation algorithm since each preprocessing step often involves averaging and/or interpolation steps that can bias SNR estimates). The first volume of the motion-corrected rest scan was visually inspected and 10 voxels outside of the head were selected. Due to the unavailability of phantom scans, we used these voxels to estimate the scanner noise, while the two have been shown to yield consistent noise estimates [73]. For each of the 10 voxels, we calculated the temporal variance of the corresponding time series and averaged the results, providing an estimate of scanner noise variance $\sigma_N^2$. To estimate the signal power, a gray matter mask was extracted using each subject's T1 scan and linearly registered back to the subject's motion corrected rest scan. We then computed the temporal variance of each gray matter voxel and averaged the results, yielding an estimate of the combined signal and noise variance. Assuming statistical independence between scanner noise and the subjects' BOLD activity, this combined variance is precisely the sum $\sigma_S^2 + \sigma_N^2$ of signal variance and noise variance. The SNR was then calculated as $\sigma_S/\sigma_N$. Note that this process is inherently conservative and provides an upper bound on the SNR, as it, for instance, does not include any physiological signals into "noise". Therefore, the ratio between the power of signals of neural origin over all other signals contributing to rsfMRI time series may be much lower than 6.5. An SNR of about 6.5, however, is still low enough to yield a significant linearizing effect, highlighting the importance of measurement noise in downstream computational modeling.

# Acknowledgements

E.N. and D.S.B. acknowledge support from the Swartz Foundation and from the National Science Foundation award #1926757. J.S. acknowledges support from the National Institute of Mental Health award #1F31MH120925-01A1. E.J.C. acknowledges support from the National Institute of Mental Health award #F30 MH118871-01. D.S.B. also acknowledges support from the John D. and Catherine T. MacArthur Foundation, the Institute for Scientific Interchange Foundation, the Paul G. Allen Family Foundation, and the Army Research Office (W911NF-16-1-0474). The content is solely the responsibility of the authors and does not necessarily represent the official views of any of the funding agencies.

## Author Contributions

EN, GJP, and DSB designed the research; EN performed the research; MAB preprocessed the fMRI data; JS preprocessed the iEEG data; MAB, JS, LC, EJC, XH, and AM assisted in the neuroimaging analysis of the data; all authors contributed in writing the paper.

## Additional Information

# References

[1] N. Kriegeskorte and P. K. Douglas, "Cognitive computational neuroscience," *Nat Neurosci*, vol. 21, no. 9, pp. 1148–1160, 2018.

[2] R. C. Wilson and Y. Niv, "Is model fitting necessary for model-based fMRI?" *PLoS Comput Biol*, vol. 11, no. 6, p. e1004237, 2015.

[3] D. A. Ruff, A. M. Ni, and M. R. Cohen, "Cognition as a window into neuronal population space," *Annu Rev Neurosci*, vol. 41, pp. 77–97, 2018.

[4] K. Amunts, C. Ebell, J. Muller, M. Telefont, A. Knoll, and T. Lippert, "The human brain project: creating a european research infrastructure to decode the human brain," *Neuron*, vol. 92, no. 3, pp. 574–581, 2016.

[5] S. Gu, F. Pasqualetti, M. Cieslak, Q. K. Telesford, B. Y. Alfred, A. E. Kahn, J. D. Medaglia, J. M. Vettel, M. B. Miller, S. T. Grafton, and D. S. Bassett, "Controllability of structural brain networks," *Nature communications*, vol. 6, no. 1, pp. 1–10, 2015.

[6] O. G. Sani, Y. Yang, M. B. Lee, H. E. Dawes, E. F. Chang, and M. M. Shanechi, "Mood variations decoded from multi-site intracranial human brain activity," *Nature biotechnology*, vol. 36, no. 10, pp. 954–961, 2018.

[7] E. M. Izhikevich, *Dynamical Systems in Neuroscience*, ser. Computational neuroscience Dynamical systems in neuroscience. MIT Press, 2007. [Online]. Available: https://books.google.com/books?id=kVjM6DFk-twC

[8] V. Booth and J. Rinzel, "A minimal, compartmental model for a dendritic origin of bistability of motoneuron firing patterns," *J Comput Neurosci*, vol. 2, no. 4, pp. 299–312, 1995.

[9] W. J. Freeman, "Nonlinear gain mediating cortical stimulus-response relations," *Biological Cybernetics*, vol. 33, no. 4, pp. 237–247, 1979.

[10] H. R. Wilson and J. D. Cowan, "Excitatory and inhibitory interactions in localized populations of model neurons," *Biophysical journal*, vol. 12, no. 1, pp. 1–24, 1972.

[11] X. Li, D. Coyle, L. Maguire, T. M. McGinnity, and H. Benali, "A model selection method for nonlinear system identification based fmri effective connectivity analysis," *IEEE transactions on medical imaging*, vol. 30, no. 7, pp. 1365–1380, 2011.

[12] Y. M. Wang, R. T. Schultz, R. T. Constable, and L. H. Staib, "Nonlinear estimation and modeling of fmri data using spatio-temporal support vector regression," in *Biennial International Conference on Information Processing in Medical Imaging*. Springer, 2003, pp. 647–659.

[13] K. E. Stephan, L. Kasper, L. M. Harrison, J. Daunizeau, H. E. den Ouden, M. Breakspear, and K. J. Friston, "Nonlinear dynamic causal models for fmri," *Neuroimage*, vol. 42, no. 2, pp. 649–662, 2008.

[14] P. Ritter, M. Schirner, A. R. McIntosh, and V. K. Jirsa, "The virtual brain integrates computational modeling and multimodal neuroimaging," *Brain Connect*, vol. 3, no. 2, pp. 121–145, 2013.

[15] C. J. Stam, "Nonlinear dynamical analysis of eeg and meg: review of an emerging field," *Clinical neurophysiology*, vol. 116, no. 10, pp. 2266–2301, 2005.

[16] C. L. Ehlers, J. Havstad, D. Prichard, and J. Theiler, "Low doses of ethanol reduce evidence for nonlinear structure in brain activity," *Journal of Neuroscience*, vol. 18, no. 18, pp. 7474–7486, 1998.

[17] E. Gultepe and B. He, "A linear/nonlinear characterization of resting state brain networks in fmri time series," *Brain topography*, vol. 26, no. 1, pp. 39–49, 2013.

[18] K. J. Blinowska and M. Malinowski, "Non-linear and linear forecasting of the eeg time series," *Biological cybernetics*, vol. 66, no. 2, pp. 159–165, 1991.

[19] Y. Zhao, S. A. Billings, H.-L. Wei, and P. G. Sarrigiannis, "A parametric method to measure time-varying linear and nonlinear causality with applications to eeg data," *IEEE Transactions on Biomedical Engineering*, vol. 60, no. 11, pp. 3141–3148, 2013.

[20] Y. Yang, O. G. Sani, E. F. Chang, and M. M. Shanechi, "Dynamic network modeling and dimensionality reduction for human ecog activity," *Journal of neural engineering*, vol. 16, no. 5, p. 056014, 2019.

[21] L. Ljung, "System identification: theory for the user," *PTR Prentice Hall, Upper Saddle River, NJ*, pp. 1–14, 1999.

[22] E. M. Izhikevich, "Simple model of spiking neurons," *IEEE Transactions on neural networks*, vol. 14, no. 6, pp. 1569–1572, 2003.

[23] C. Gorrostieta, M. Fiecas, H. Ombao, E. Burke, and S. Cramer, "Hierarchical vector auto-regressive models and their applications to multi-subject effective connectivity," *Frontiers in computational neuroscience*, vol. 7, p. 159, 2013.

[24] S. A. Kim and S. Ching, "Quasilinearization-based controllability analysis of neuronal rate networks," in *2016 American Control Conference (ACC)*. IEEE, 2016, pp. 7371–7376.

[25] G. Buzsáki, C. A. Anastassiou, and C. Koch, "The origin of extracellular fields and currents—eeg, ecog, lfp and spikes," *Nature reviews neuroscience*, vol. 13, no. 6, pp. 407–420, 2012.

[26] D. N. Greve, G. G. Brown, B. A. Mueller, G. Glover, and T. T. Liu, "A survey of the sources of noise in fmri," *Psychometrika*, vol. 78, no. 3, pp. 396–416, 2013.

[27] Y. Liu, W. Coon, A. De Pesters, P. Brunner, and G. Schalk, "The effects of spatial filtering and artifacts on electrocorticographic signals," *Journal of neural engineering*, vol. 12, no. 5, p. 056008, 2015.

[28] Z. Yang, "Incorporating structural bias into neuralnetworks for natural language processing," Ph.D. dissertation, Carnegie Mellon University, 2019.

[29] A. V. Kononova, D. W. Corne, P. De Wilde, V. Shneer, and F. Caraffini, "Structural bias in population-based algorithms," *Information Sciences*, vol. 298, pp. 468–490, 2015.

[30] R. Mehta, C. Shen, T. Xu, and J. T. Vogelstein, "A consistent independence test for multivariate timeseries," *arXiv preprint arXiv:1908.06486*, 2019.

[31] M. P. Dafilis, N. C. Sinclair, P. J. Cadusch, and D. T. Liley, "Re-evaluating the performance of the nonlinear prediction error for the detection of deterministic dynamics," *Physica D: Nonlinear Phenomena*, vol. 240, no. 8, pp. 695–700, 2011.

[32] T. Deneux and O. Faugeras, "Using nonlinear models in fmri data analysis: model selection and activation detection," *NeuroImage*, vol. 32, no. 4, pp. 1669–1689, 2006.

[33] Z. Liu, C. Rios, N. Zhang, L. Yang, W. Chen, and B. He, "Linear and nonlinear relationships between visual stimuli, eeg and bold fmri signals," *Neuroimage*, vol. 50, no. 3, pp. 1054–1066, 2010.

[34] P. Wobst, R. Wenzel, M. Kohl, H. Obrig, and A. Villringer, "Linear aspects of changes in deoxygenated hemoglobin concentration and cytochrome oxidase oxidation during brain activation," *Neuroimage*, vol. 13, no. 3, pp. 520–530, 2001.

[35] O. I. Rumyantsev, J. A. Lecoq, O. Hernandez, Y. Zhang, J. Savall, R. Chrapkiewicz, J. Li, H. Zeng, S. Ganguli, and M. J. Schnitzer, "Fundamental bounds on the fidelity of sensory cortical coding," *Nature*, vol. 580, no. 7801, pp. 100–105, 2020.

[36] H. K. Khalil, *Nonlinear Systems*, ser. Pearson Education. Prentice Hall, 2002. [Online]. Available: https://books.google.com/books?id=t_d1QgAACAAJ

[37] J. M. Palva and S. Palva, "Functional integration across oscillation frequencies by cross-frequency phase synchronization," *Eur J Neurosci*, vol. 48, no. 7, pp. 2399–2406, 2018.

[38] J. G. T. Zañudo, G. Yang, and R. Albert, "Structure-based control of complex networks with nonlinear dynamics," *Proc Natl Acad Sci U S A*, vol. 114, no. 28, pp. 7234–7239, 2017.

[39] J. C. Rozum and R. Albert, "Identifying (un)controllable dynamical behavior in complex networks," *PLoS Comput Bio*, vol. 14, no. 12, p. e1006630, 2018.

[40] E. Tang and D. S. Bassett, "Colloquium: Control of dynamics in brain networks," *Reviews of modern physics*, vol. 90, no. 3, p. 031003, 2018.

[41] E. K. Towlson, P. E. Vértes, G. Yan, Y. L. Chew, D. S. Walker, W. R. Schafer, and A. L. Barabási, "Caenorhabditis elegans and the network control framework-FAQs," *Philos Trans R Soc Lond B Biol Sci*, vol. 373, no. 1758, p. 20170372, 2018.

[42] T. M. Karrer, J. Z. Kim, J. Stiso, A. E. Kahn, F. Pasqualetti, U. Habel, and D. S. Bassett, "A practical guide to methodological considerations in the controllability of structural brain networks," *J Neural Eng*, vol. 17, no. 2, p. 026031, 2020.

[43] C. Goutte, F. A. Nielsen, and L. K. Hansen, "Modeling the hemodynamic response in fmri using smooth fir filters," *IEEE transactions on medical imaging*, vol. 19, no. 12, pp. 1188–1201, 2000.

[44] M. Haller, T. Donoghue, E. Peterson, P. Varma, P. Sebastian, R. Gao, T. Noto, R. T. Knight, A. Shestyuk, and B. Voytek, "Parameterizing neural power spectra," *BioRxiv*, p. 299859, 2018.

[45] K. Bansal, J. Nakuci, and S. F. Muldoon, "Personalized brain network models for assessing structure-function relationships," *Curr Opin Neurobiol*, vol. 52, pp. 42–47, 2018.

[46] M. Schirner, S. Rothmeier, V. K. Jirsa, A. R. McIntosh, and P. Ritter, "An automated pipeline for constructing personalized virtual brains from multimodal neuroimaging data," *Neuroimage*, vol. 117, pp. 343–357, 2015.

[47] S. Bayrak, P. Hövel, and V. Vuksanović, "Modeling functional connectivity on empirical and randomized structural brain networks," in *Differ Equ Dyn Syst*. Springer, 2017.

[48] V. M. Saenger, J. Kahan, T. Foltynie, K. Friston, T. Z. Aziz, A. L. Green, T. J. van Hartevelt, J. Cabral, A. B. Stevner, H. M. Fernandes, *et al.*, "Uncovering the underlying mechanisms and whole-brain dynamics of deep brain stimulation for parkinson's disease," *Scientific reports*, vol. 7, no. 1, pp. 1–14, 2017.

[49] T. S. Zarghami and K. J. Friston, "Dynamic effective connectivity," *Neuroimage*, vol. 207, p. 116453, 2020.

[50] K. J. Friston, A. Bastos, V. Litvak, K. E. Stephan, P. Fries, and R. J. Moran, "DCM for complex-valued data: cross-spectra, coherence and phase-delays," *Neuroimage*, vol. 59, no. 1, pp. 439–455, 2012.

[51] C. O. Becker, D. S. Bassett, and V. M. Preciado, "Large-scale dynamic modeling of task-fmri signals via subspace system identification," *Journal of neural engineering*, vol. 15, no. 6, p. 066016, 2018.

[52] Y. Yang, A. T. Connolly, and M. M. Shanechi, "A control-theoretic system identification framework and a real-time closed-loop clinical simulation testbed for electrical brain stimulation," *J Neural Eng*, vol. 15, no. 6, p. 066007, 2018.

[53] D. M. Barch, "Resting-state functional connectivity in the human connectome project: Current status and relevance to understanding psychopathology," *Harv Rev Psychiatry*, vol. 25, no. 5, pp. 209–217, 2017.

[54] G. C. Burgess, S. Kandala, D. Nolan, T. O. Laumann, J. D. Power, B. Adeyemo, M. P. Harms, S. E. Petersen, and D. M. Barch, "Evaluation of denoising strategies to address motion-correlated artifacts in resting-state functional magnetic resonance imaging data from the human connectome project," *Brain Connect*, vol. 6, no. 9, pp. 669–680, 2016.

[55] J. Elam, "Hcp data release updates: Known issues and planned fixes," https://wiki.humanconnectome.org/display/PublicData/HCP+Data+Release+Updates%3A+Known+Issues+and+Planned+fixes, May 2020, last accessed May 13, 2020.

[56] A. Schaefer, R. Kong, E. M. Gordon, T. O. Laumann, X.-N. Zuo, A. J. Holmes, S. B. Eickhoff, and B. T. Yeo, "Local-global parcellation of the human

cerebral cortex from intrinsic functional connectivity mri," *Cerebral Cortex*, vol. 28, no. 9, pp. 3095–3114, 2018.

[57] Y. Tian, D. S. Margulies, M. Breakspear, and A. Zalesky, "Hierarchical organization of the human subcortex unveiled with functional connectivity gradients," *bioRxiv*, 2020.

[58] J. Stiso, A. N. Khambhati, T. Menara, A. E. Kahn, J. M. Stein, S. R. Das, R. Gorniak, J. Tracy, B. Litt, K. A. Davis, F. Pasqualetti, T. H. Lucas, and D. S. Bassett, "White matter network architecture guides direct electrical stimulation through optimal state transitions," *Cell Rep*, vol. 28, no. 10, pp. 2554–2566.e7, 2019.

[59] A. N. Khambhati, A. E. Kahn, J. Costantini, Y. Ezzyat, E. A. Solomon, R. E. Gross, B. C. Jobst, S. A. Sheth, K. A. Zaghloul, G. Worrell, S. Seger, B. C. Lega, S. Weiss, M. R. Sperling, R. Gorniak, S. R. Das, J. M. Stein, D. S. Rizzuto, M. J. Kahana, T. H. Lucas, K. A. Davis, J. I. Tracy, and D. S. Bassett, "Functional control of electrophysiological network architecture using direct neurostimulation in humans," *Netw Neurosci*, vol. 3, no. 3, pp. 848–877, 2019.

[60] R. F. Betzel, J. D. Medaglia, A. E. Kahn, J. Soffer, D. R. Schonhaut, and D. S. Bassett, "Structural, geometric and genetic factors predict interregional brain connectivity patterns probed by electrocorticography," *Nat Biomed Eng*, vol. 3, no. 11, pp. 902–916, 2019.

[61] V. Lawhern, W. D. Hairston, K. McDowell, M. Westerfield, and K. Robbins, "Detection and classification of subject-generated artifacts in eeg signals using autoregressive models," *Journal of neuroscience methods*, vol. 208, no. 2, pp. 181–189, 2012.

[62] S. L. Bressler, C. G. Richter, Y. Chen, and M. Ding, "Cortical functional network organization from autoregressive modeling of local field potential oscillations," *Statistics in medicine*, vol. 26, no. 21, pp. 3875–3885, 2007.

[63] R. Deshpande, G.-R. Wu, D. Marinazzo, X. Hu, and G. Deshpande, "Hemodynamic response function (hrf) variability confounds resting-state fmri functional connectivity," *Magnetic resonance in medicine*, vol. 80, no. 4, pp. 1697–1713, 2018.

[64] A. J. Taylor, J. H. Kim, and D. Ress, "Characterization of the hemodynamic response function across the majority of human cerebral cortex," *NeuroImage*, vol. 173, pp. 322–331, 2018.

[65] M. Singh, T. Braver, M. Cole, and S. Ching, "Individualized dynamic brain models: Estimation and validation with resting-state fmri," *bioRxiv*, p. 678243, 2019.

[66] J. Roll, "Local and piecewise affine approaches to system identification," Ph.D. dissertation, Linkoping University, 2003.

[67] L. Ljung, "Approaches to identification of nonlinear systems," in *Proceedings of the 29th Chinese Control Conference*. IEEE, 2010, pp. 1–5.

[68] D. Popivanov, J. Dushanova, A. Mineva, and I. Krekule, "Detection of successive changes in dynamics of eeg time series: linear and nonlinear approach," in *Proceedings of 18th Annual International Conference of the IEEE Engineering in Medicine and Biology Society*, vol. 4. IEEE, 1996, pp. 1590–1591.

[69] H. V. Poor, *An introduction to signal detection and estimation*. Springer Science & Business Media, 2013.

[70] G. E. Box and D. A. Pierce, "Distribution of residual autocorrelations in autoregressive-integrated moving average time series models," *Journal of the American statistical Association*, vol. 65, no. 332, pp. 1509–1526, 1970.

[71] G. M. Ljung and G. E. P. Box, "On a measure of lack of fit in time series models," *Biometrika*, vol. 65, no. 2, pp. 297–303, 1978.

[72] W. K. Li and A. I. McLeod, "Distribution of the residual autocorrelations in multivariate arma time series models," *Journal of the Royal Statistical Society: Series B (Methodological)*, vol. 43, no. 2, pp. 231–239, 1981.

[73] C.-C. Chen and C. W. Tyler, "Spectral analysis of fmri signal and noise," in *Novel trends in brain science*. Springer, 2008, pp. 63–76.

[74] S. M. Mitchell, S. Lange, and H. Brus, "Gendered citation patterns in international relations journals," *International Studies Perspectives*, vol. 14, no. 4, pp. 485–492, 2013.

[75] M. L. Dion, J. L. Sumner, and S. M. Mitchell, "Gendered citation patterns across political science and social science methodology fields," *Political Analysis*, vol. 26, no. 3, pp. 312–327, 2018.

[76] N. Caplar, S. Tacchella, and S. Birrer, "Quantitative evaluation of gender bias in astronomical publications from citation counts," *Nature Astronomy*, vol. 1, no. 6, p. 0141, 2017.

[77] D. Maliniak, R. Powers, and B. F. Walter, "The gender citation gap in international relations," *International Organization*, vol. 67, no. 4, pp. 889–922, 2013.

[78] J. D. Dworkin, K. A. Linn, E. G. Teich, P. Zurn, R. T. Shinohara, and D. S. Bassett,

"The extent and drivers of gender imbalance in neuroscience reference lists," *bioRxiv*, 2020. [Online]. Available: https://www.biorxiv.org/content/early/2020/01/11/2020.01.03.894378

[79] D. Zhou, E. J. Cornblath, J. Stiso, E. G. Teich, J. D. Dworkin, A. S. Blevins, and D. S. Bassett, "Gender diversity statement and code notebook v1.0," Feb. 2020. [Online]. Available: https://doi.org/10.5281/zenodo.3672110

[80] A. Ambekar, C. Ward, J. Mohammed, S. Male, and S. Skiena, "Name-ethnicity classification from open sources," in *Proceedings of the 15th ACM SIGKDD international conference on Knowledge Discovery and Data Mining*, 2009, pp. 49–58.

[81] G. Sood and S. Laohaprapanon, "Predicting race and ethnicity from the sequence of characters in a name," *arXiv preprint arXiv:1805.02109*, 2018.